\documentclass[prd,preprint,superscriptaddress,preprintnumbers,eqsecnum,showpacs,nofootinbib,nobibnotes,noeprint]{revtex4-1}
\usepackage[linkcolor=blue,citecolor=blue,urlcolor=blue,colorlinks=true,breaklinks]{hyperref} 

\usepackage{amsfonts,bm}
\usepackage{amsfonts,amssymb,amsmath}
\usepackage{bbold}
\usepackage{graphicx}
\usepackage{enumitem}
\usepackage[english]{babel}
\usepackage{slashed}
\usepackage[usenames]{xcolor}
\usepackage{mathtools}
\usepackage[makeroom]{cancel}
\usepackage{color}

\def\srm#1{{\rm{\scriptscriptstyle #1}}}

\newcommand{\be}{\begin{equation}}
\newcommand{\bea}{\begin{eqnarray}}
\newcommand{\ee}{\end{equation}}
\newcommand{\eea}{\end{eqnarray}}

\def\asym{\srm{asym}}
\newcommand{\Zfourasym}{Z_{4}^\asym}
\newcommand{\Zthreeasym}{Z_{3}^\asym}
\newcommand{\Zoneasym}{Z_{1}^\asym}

\newcommand{\Zgasym}{Z_{g}^\asym}
\newcommand{\alphaasym}{\alpha_{s}^\asym}

\def\1eq#1{Eq.~(\ref{#1})}

\def\2eqs#1#2{Eqs.~(\ref{#1}) and~(\ref{#2})}
\def\3eqs#1#2#3{Eqs.~(\ref{#1}),~(\ref{#2}) and~(\ref{#3})}

\def\fig#1{Fig.~\ref{#1}}

\def\ie{{\it i.e.}, }

\def\s#1{{\scriptscriptstyle #1}}

\def\g{\Gamma}

\def\gbar{{\overline \Gamma}\vphantom{\Gamma} }
\def\T{ \g }

\newcommand{\fatg}{{\rm{I}}\!\Gamma}

\newcommand{\Ls}{ \mathit{L}_{{sg}}}

\newcommand{\Luv}{ \mathit{L}_{\srm{UV}}}

\def\sb{\bar{s}}

\def\T{  G }

\def\deg{G^{*}_1}

\def\degx#1{G^{*#1}_1}   

\def\LambdaMOM{\Lambda_{\srm{MOM}}}

\def\LambdaUV{\Lambda_{\srm{UV}}}

\begin{document}

\title{Four-gluon vertex in collinear kinematics}

\author{A.~C.~Aguilar}
\affiliation{\mbox{University of Campinas - UNICAMP, Institute of Physics Gleb Wataghin,} \\
13083-859 Campinas, S\~{a}o Paulo, Brazil}

\author{M.~N.~Ferreira}
\affiliation{\mbox{Department of Theoretical Physics and IFIC, 
University of Valencia and CSIC},
E-46100, Valencia, Spain}

\author{J.~Papavassiliou}
\affiliation{\mbox{Department of Theoretical Physics and IFIC, 
University of Valencia and CSIC},
E-46100, Valencia, Spain}

\author{L.~R.~Santos}
\affiliation{\mbox{University of Campinas - UNICAMP, Institute of Physics Gleb Wataghin,} \\
13083-859 Campinas, S\~{a}o Paulo, Brazil}

\begin{abstract}
To date, the four-gluon vertex is the least explored 
component of the QCD Lagrangian, mainly due to 
the vast proliferation of Lorentz and color 
structures required for its description. 
In this work we present a 
nonperturbative 
study of this vertex, based on  
the one-loop dressed 
Schwinger-Dyson equation obtained from the 
4PI effective action. 
A vast simplification 
is brought about by 
resorting to 
``collinear'' kinematics, where all 
momenta are parallel to each other,
and by appealing to the 
charge conjugation symmetry in order to eliminate certain color structures. Out of the 
fifteen form factors that 
comprise the 
transversely-projected version 
of this vertex, two are singled out and studied 
in detail;  the one 
associated with the 
classical tensorial structure 
is moderately suppressed  
in the infrared regime, 
while the other 
diverges logarithmically at 
the origin. 
Quite interestingly, both 
form factors display the 
property known as 
``planar degeneracy'' at a rather 
high level of accuracy. 
With these results we 
construct an effective charge 
that quantifies the strength of the four-gluon interaction, and compare it 
with other 
vertex-derived 
charges from the 
gauge sector of QCD.

\end{abstract}

\maketitle

\section{Introduction}
\label{intro}

In recent years, our understanding of 
nonperturbative QCD has advanced considerably 
thanks to the systematic scrutiny of  
the fundamental correlation functions (Green's functions) 
of the theory. This ongoing exploration
proceeds through continuous studies 
based on functional methods, such as 
Schwinger-Dyson equations (SDEs)~\cite{Roberts:1994dr,Alkofer:2000wg,Fischer:2006ub,Roberts:2007ji,Binosi:2009qm,Cloet:2013jya,Aguilar:2015bud,Huber:2018ned,Papavassiliou:2022wrb,Binosi:2014aea,Maris:1997tm, Maris:2003vk,Eichmann:2008ef, Cloet:2008re,Boucaud:2008ky,Eichmann:2009qa,Fischer:2008uz,Boucaud:2008ky,RodriguezQuintero:2010wy,Pennington:2011xs,Huber:2012zj,Cloet:2013jya,Boucaud:2008ky, Roberts:2020hiw,Roberts:2020hiw,Gao:2021wun, Huber:2016tvc,  Huber:2020keu} or the function
renormalization group~\cite{Cyrol:2014kca,Braun:2007bx,Fister:2013bh,Pawlowski:2003hq,Pawlowski:2005xe,Cyrol:2017ewj,Cyrol:2018xeq,Corell:2018yil,Blaizot:2021ikl,Horak:2021pfr}\footnote{For related studies in the context of 
the Curci-Ferrari model, see~\cite{Tissier:2010ts,Dudal:2008sp,Mintz:2017qri,Barrios:2020ubx}.},
and by means of 
gauge-fixed lattice 
simulations~\cite{Cucchieri:2006tf,Cucchieri:2007md,Bogolubsky:2007ud,Bogolubsky:2009dc,Cucchieri:2008qm,Cucchieri:2009zt, Oliveira:2009eh,Oliveira:2010xc,Ayala:2012pb,Oliveira:2012eh,Athenodorou:2016oyh,Duarte:2016ieu,
Sternbeck:2017ntv,Boucaud:2017ksi,Boucaud:2018xup,Aguilar:2021okw,Maas:2011se}.
In particular, a great deal of information  
is available on the two-point sector of QCD
(propagators of the gluon, ghost, and quark fields), 
as well as the three-point sector (ghost-gluon, three-gluon, 
and quark-gluon vertices); 
for related reviews see~\mbox{\cite{Roberts:1994dr,Alkofer:2000wg,Pawlowski:2005xe,Fischer:2006ub,Binosi:2009qm,
Roberts:2007ji,Cloet:2013jya,
Maas:2011se,Aguilar:2015bud,Huber:2018ned,Dupuis:2020fhh,Ding:2022ows,Papavassiliou:2022wrb,Ferreira:2023fva}}.
This knowledge, in turn, enables 
the reliable determination of 
observables built out of these functions, and 
allows for the detailed scruting of 
underlying physical mechanisms, associated 
with the 
emergence of fundamental mass scales, 
formation of bound states, and confinement~\mbox{\cite{Cornwall:1981zr,Aguilar:2004sw,Aguilar:2002tc, Aguilar:2008xm,Binosi:2009qm,Cornwall:2010upa, Binosi:2012sj,Ibanez:2012zk,Aguilar:2015bud,Aguilar:2016vin,Binosi:2017rwj,Aguilar:2017dco,Gao:2017uox,Boucaud:2011ug,Vandersickel:2012tz,Cloet:2013jya,Meyer:2015eta,Kondo:2014sta,Aguilar:2015bud,Eichmann:2016yit,Cyrol:2017ewj,Huber:2018ned,Greensite:2003bk}}.

Instead, the nonperturbative aspects of the four-gluon vertex, 
to be denoted by $\fatg^{abcd}_{\mu\nu\rho\sigma}$,
are relatively poorly known~\cite{Driesen:1998xc,Kellermann:2008iw,Binosi:2014kka,Cyrol:2014kca,Huber:2016tvc,Huber:2017txg, Huber:2020keu,Pawlowski:2022zhh}; for perturbative studies, see~\cite{Pascual:1980yu,Brandt:1985zz, Papavassiliou:1992ia, Hashimoto:1994ct, Gracey:2014ola, Gracey:2017yfi,Ahmadiniaz:2013rla,Ahmadiniaz:2016qwn}. 
The main reason for this limitation is the 
large proliferation of Lorentz and color structures, 
which makes the treatment of this vertex exceedingly 
cumbersome in the continuum, and  overly  
costly on the lattice. 
Since this vertex is inextricably 
connected with all other 
Green's functions through the 
coupled functional equations 
that govern their evolution, 
it is certainly desirable to improve our 
understanding of its dynamics. 

Certain new insights gained 
from recent studies of the three-gluon vertex
appear particularly promising  
for determining the leading nonperturbative features 
of the four-gluon vertex.   
In particular, the logistics of the three-gluon vertex are greatly simplified 
due to the property known as 
``planar degeneracy'': with an error of 
less than 10\%, 
the form factor associated 
with the classical tensor 
is the same for all momentum configurations 
lying on a given plane~\cite{Pinto-Gomez:2022brg,Aguilar:2023qqd,Ferreira:2023fva}. 
Even though the dynamical reason that 
enforces this feature is not fully understood, 
its manifestation hinges crucially on the 
appropriate exploitation of the 
Bose symmetry of the three-gluon vertex, 
a symmetry shared also 
by the four-gluon vertex. 
It is, therefore, possible that 
the planar degeneracy may be 
a property of the four-gluon vertex, leading to a 
great simplification 
of practical computations.    
Parallel to these developments, lattice simulations 
have been carrying out exploratory studies of 
the four-gluon vertex in 
simplified kinematics~\cite{Catumba:2021qbh,Colaco:2023qin,Colaco:2024gmt}, and are expected to 
access a wider array of momentum configurations
in the near future.

In the present work we study the quenched four-gluon vertex 
(no dynamical quarks) in the context of the one-loop SDE  
derived from the four-particle irreducible (4PI) effective 
action~\cite{Cornwall:1974vz,Cornwall:1973ts,Berges:2004pu,York:2012ib,Williams:2015cvx}  at four loops~\cite{Carrington:2010qq,Carrington:2013koa}. Note that, within this formalism, 
the extremization of the $n$-loop $n$-PI effective action with respect to the 
$m$-point correlation functions
($m \leq n$) 
generates the corresponding equations of motions (SDEs),  which 
are expressed in terms of  \mbox{$(n-m+1)$-loop} diagrams~\cite{Berges:2004pu,Carrington:2010qq,Carrington:2013koa}. We emphasize that our analysis 
does {\it not} treat the entire system of coupled SDEs that govern the propagators and vertices 
entering in the SDE of the four-gluon vertex: 
instead, we consider the four-gluon SDE in isolation, 
using lattice results as inputs for all other 
dynamical components.

The focal point of our attention is the \emph{transversely projected}
four-gluon vertex, 
$\overline{\fatg}^{abcd}_{\mu\nu\rho\sigma}$,
which is precisely the one   
simulated on the lattice in the Landau gauge~\cite{Catumba:2021qbh,Colaco:2023qin,Colaco:2024gmt}, and appears in the majority of physical applications~\cite{Fukamachi:2016wxf,Huber:2021yfy,Pawlowski:2022zhh}. The analysis is restricted to the 
case of the  ``collinear'' kinematics, 
where all 
vertex momenta are 
parallel to each other, or, equivalently, 
proportional to a single 
momentum $p$ (\ie $p_i=x_i p$).  
This kinematic choice, 
together with the proper exploitation 
of the charge conjugation symmetry, 
drastically reduces 
the tensorial structure of  $\overline{\fatg}^{abcd}_{\mu\nu\rho\sigma}$: only 
15 elements, $t_{i,\mu\nu\rho\sigma}^{abcd}$, 
are required for the full description of its 
Lorentz and color content, 
with the associated form factors denoted by $\T_{i}$ ($i=1,\ldots,15$). Note that this basis 
contains as one of its elements 
the transversely 
projected classical (tree-level) 
tensor of the four-gluon vertex; in particular, 
\mbox{$t_1 = 
\overline{\Gamma}_{\!\!0}$}. 

There is an additional simplification that we will 
incorporate in our study, 
prompted by the following two  
observations. First, 
the elements 
$t_1$, $t_2$, and $t_3$
satisfy a special orthogonality condition  
with respect to the rest of the basis. Second, they depend only on $p$, being independent of 
the variables 
$x_i$. These facts reduce 
significantly the algebraic complexity of the problem, facilitating the determination of the attendant  
form factors $\T_{1}, \T_{2}$,  and $\T_{3}$; we therefore focus 
exclusively on this special 
subset of contributions. 

In the 4PI formalism, 
all propagators and vertices 
appearing in the 
one-loop diagrams 
comprising the SDE 
that governs the 
evolution of the $G_i$ 
are fully dressed, 
including the four-gluon vertices. 
This converts the SDE into 
an integral equation, involving the 
unknown function 
$\overline{\fatg}^{abcd}_{\mu\nu\rho\sigma}$ evaluated in different kinematic configurations.  
The approximation used for 
dealing with this 
complication 
is to assume that 
the tree-level form factor 
$G_1$ is dominant, and 
that it satisfies 
planar degeneracy, in exact 
analogy to 
its three-gluon vertex counterpart. 

The results of our study may be summarized as follows:

({\it i}) 
Under the approximations 
mentioned above, the form factor $\T_{2}$ vanishes identically. 

({\it ii})  $\T_{1}$ and $\T_{3}$ 
are computed for 
several collinear configurations. We find that $\T_{1}$ is 
suppressed in the infrared with respect to its tree-level value, 
$G_1^{(0)} =1$, 
reaching a finite value at the origin, while 
$\T_{3}$ displays  
a logarithmic divergence in the deep infrared,  originating from the ghost loops.

({\it iii})  
For momenta below $1$ GeV, 
our results for $\T_{1}$ are compatible with planar degeneracy to within 1.7$\%$.  In the case of 
$G_3$, a similar tendency is 
observed, albeit with a lesser 
degree of accuracy.

({\it iv}) 
With the help of the $\T_{1}$ 
computed 
for an extensive set of collinear configurations,
we construct a ``band'' of effective charges, $\alpha_{\srm{4g}}(p)$, which 
may be used to 
quantify the strength of the four-gluon interaction. The 
$\alpha_{\srm{4g}}(p)$
is then 
compared with 
the effective charges 
$\alpha_{\srm{3g}}(p)$ and 
$\alpha_{\srm{cg}}(p)$,
obtained from the soft-gluon limits of the three- and ghost-gluon vertices, respectively.
For momenta below $1.5$ GeV, we find the clear hierarchy 
\mbox{$\alpha_{\srm{3g}}(p) < \alpha_{\srm{4g}}(p) < 
\alpha_{\srm{cg}}(p)$}, 
in agreement with earlier results 
presented in~\cite{Cyrol:2014kca,Huber:2018ned}.

The article is organized as follows. 
In Sec.~\ref{sec:gen}  we summarize the general features of $\fatg^{abcd}_{\mu\nu\rho\sigma}$ in collinear configurations.  In Sec.~\ref{sec:color} we show how the  charge conjugation symmetry prohibits the
presence of color tensors of the type $f^{abx} d^{cdx}$.   
In Sec.~\ref{sec:basis} we define the Lorentz tensor basis for collinear configurations, and  
determine the projectors  
that allow us to extract the 
attendant form factors.
In Sec.~\ref{sec:sde}  we present the SDE governing the evolution of $\fatg^{abcd}_{\mu\nu\rho\sigma}$, derived from the 4PI effective action  at the four-loop level, and its renormalization. 
In Sec.~\ref{sec:num}, the SDE is solved for general collinear configurations, 
and the form factors $G_{1,3}$
are determined and analyzed. In Section~\ref{sec:comparison} we compare our results to previous continuum studies, and comment on the impact of the three-gluon vertex in the four-gluon SDE. In 
Sec.~\ref{sec:Gi_finite} we present an analytic 
proof of the ultraviolet finiteness of 
the form factors $G_{1,3}$, which complements our 
numerical findings.
Next, in Sec.~\ref{sec:charge},  we compute the four-gluon effective charge.  Finally, our conclusions are discussed in Sec.~\ref{conc},
and certain technical points are summarized in two 
Appendices.


\section{Four-gluon vertex in collinear kinematics}
\label{sec:gen}

\begin{figure}[t!]
    \centering
\includegraphics[width=0.8\linewidth]{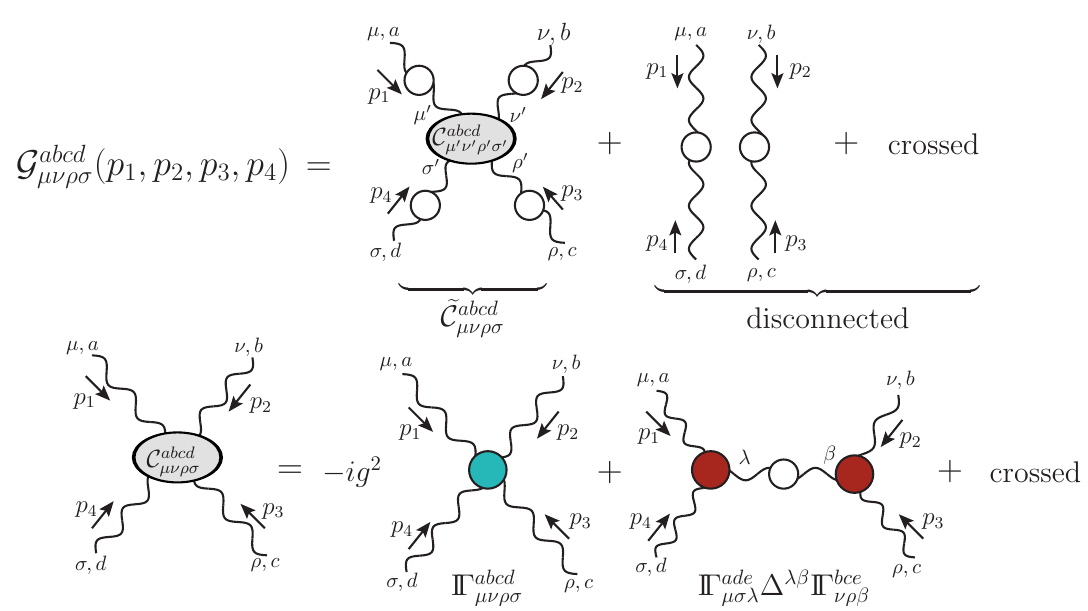}
     \caption { Upper panel:  Diagrammatic representations of the full four-gluon Green's function, $ \mathcal{G}^{abcd}_{\mu\nu\rho\sigma}(p_1,p_2,p_3,p_4)$,  separated into the connected,  $\mathcal{\widetilde C}^{abcd}_{\mu\nu\rho\sigma}$, and disconnected parts. Lower panel: Schematic decomposition of the
     \emph{amputated} four-gluon Green's function, $\mathcal{C}^{abcd}_{\mu\nu\rho\sigma}$,  into the 
1PI vertex, $-ig^2\fatg^{abcd}_{\mu\nu\rho\sigma}(p_1,p_2,p_3,p_4)$, and  the 1PR terms of the type 
$\fatg^{ade}_{\mu\sigma\lambda}
\Delta^{\lambda\beta}
\fatg^{bce}_{\nu\rho\beta}$ and 
crossed contributions.}
     \label{fig:1P1}
\end{figure}

The starting point is the definition 
of the four-gluon correlation function as the vacuum expectation value of the time-ordered product 
of four SU(3) gauge fields, $\widetilde{A}^a_\mu(p)$, in momentum space, 
\begin{align}
\mathcal{G}^{abcd}_{\mu\nu\rho\sigma}(p_1,p_2,p_3,p_4) = \langle 0 \vert T [\widetilde{A}^a_\mu(p_1) \widetilde{A}^b_\nu(p_2) \widetilde{A}^c_\rho(p_3)\widetilde{A}^d_\sigma(p_4)] \vert 0\rangle 
\,,
\label{eq:Gdef}
\end{align}
with $p_1+p_2+p_3+p_4 = 0$. The diagrammatic representation 
of $\mathcal{G}^{abcd}_{\mu\nu\rho\sigma}$, shown in the upper panel of \fig{fig:1P1}, distinguishes 
between connected and disconnected contributions.
The connected Green's function, denoted 
by $\mathcal{\widetilde C}^{abcd}_{\mu\nu\rho\sigma}$, 
has the general form 
\begin{align}
\mathcal{\widetilde C}^{abcd}_{\mu\nu\rho\sigma}(p_1,p_2,p_3,p_4)
= \Delta^{\mu'}_{\mu}(p_1) 
\Delta^{\nu'}_{\nu}(p_2)\Delta^{\rho'}_{\rho}(p_3)\Delta^{\sigma'}_{\sigma}(p_4)\,
\mathcal{C}^{abcd}_{\mu'\nu'\rho'\sigma'}(p_1,p_2,p_3,p_4)\,,
\label{eq:Ctdef}
\end{align}
where $\mathcal{C}^{abcd}_{\mu\nu\rho\sigma}$ 
is the \emph{amputated} vertex. 
$\mathcal{C}^{abcd}_{\mu\nu\rho\sigma}$ may be 
further separated into one-particle irreducible 
(1PI) and one-particle reducible (1PR) contributions,
namely
\begin{align}
\mathcal{C}^{abcd}_{\mu\nu\rho\sigma}(p_1,p_2,p_3,p_4) = 
-ig^2\underbrace{ \fatg^{abcd}_{\mu\nu\rho\sigma}(p_1,p_2,p_3,p_4)}_{\text{1PI}} 
\,-i \underbrace{\fatg^{ade}_{\mu\sigma\lambda}\Delta^{\lambda\beta}
\fatg^{bce}_{\nu\rho\beta} + \mathrm{crossed}}_{\text{1PR}}\,,
\label{eq:Cdef}
\end{align}
where \mbox{$\Delta_{\mu\nu}^{ab}(q)=-i\delta^{ab}\Delta_{\mu\nu}(q)$}  denotes the gluon propagator, and 
$\fatg^{abc}_{\alpha\beta\gamma}(q,r,p)$ 
the full three-gluon vertex [see lower panel of \fig{fig:1P1}].

In the Landau gauge, where  
\begin{align}
\Delta_{\mu\nu}(q) = \Delta(p^2) P_{\mu\nu}(q) \,, \qquad 
P_{\mu\nu}(q) = g_{\mu\nu} - q_{\mu} q_{\nu}/q^2 \,,
\label{eq:prop}
\end{align}
it is natural to define, at the level of \1eq{eq:Ctdef}, 
\begin{align}
\mathcal{\widetilde C}^{abcd}_{\mu\nu\rho\sigma}(p_1,p_2,p_3,p_4)
= \Delta(p^2_1) \Delta(p^2_2) \Delta(p^2_3)\Delta(p^2_4)\,
\mathcal{\overline C}^{abcd}_{\mu\nu\rho\sigma}(p_1,p_2,p_3,p_4) \,,
\label{eq:Ctdefproj}
\end{align}
where 
\begin{align}
\mathcal{\overline C}^{abcd}_{\mu\nu\rho\sigma}(p_1,p_2,p_3,p_4) := 
 P^{\mu'}_{\mu}(p_1) P^{\nu'}_{\nu}(p_2)P^{\rho'}_{\rho}(p_3)P^{\sigma'}_{\sigma}(p_4)\,\mathcal{C}^{abcd}_{\mu'\nu'\rho'\sigma'}(p_1,p_2,p_3,p_4)\,, 
\label{eq:Cbar}
\end{align}
is the \emph{transversely-projected}
amputated vertex. 
Substituting \1eq{eq:Cdef} into \1eq{eq:Cbar}, we 
obtain 
\begin{align}
\mathcal{\overline C}^{abcd}_{\mu\nu\rho\sigma}(p_1,p_2,p_3,p_4) = -ig^2 
\overline{\fatg}^{abcd}_{\mu\nu\rho\sigma}(p_1,p_2,p_3,p_4) -i 
\left\{\overline\fatg^{ade}_{\mu\sigma\lambda}
\Delta^{\lambda\beta}
\overline\fatg^{bce}_{\nu\rho\beta} + \mathrm{crossed}\,
\right\}\,,
\label{eq:allbars}
\end{align}
where
\begin{align}
\overline{\fatg}^{abcd}_{\mu\nu\rho\sigma}(p_1,p_2,p_3,p_4) = P^{\mu'}_\mu(p_1)P^{\nu'}_\nu(p_2)P^{\rho'}_\rho(p_3)P^{\sigma'}_\sigma(p_4)\fatg^{abcd}_{\mu'\nu'\rho'\sigma'}(p_1,p_2,p_3,p_4) \,, 
\label{4g_transv}
\end{align}
is the transversely-projected 1PI
four-gluon vertex, and 
\begin{align}
\overline\fatg^{abc}_{\alpha\beta\gamma}(q,r,p)
= P^{\alpha'}_\alpha(q)P^{\beta'}_\beta(r)P^{\gamma'}_\gamma(p)
\fatg^{abc}_{\alpha'\beta'\gamma'}(q,r,p)
\,,
\end{align}
is the transversely-projected three-gluon vertex.

In general kinematics, 
the vertices $\fatg^{abcd}_{\mu\nu\rho\sigma}$ 
and $\overline{\fatg}^{abcd}_{\mu\nu\rho\sigma}$
have a proliferation of tensorial and color structures, which give rise to a large number of 
form factors.
In the present work we restrict ourselves to the nonperturbative study of  
$\overline{\fatg}^{abcd}_{\mu\nu\rho\sigma}$
in \emph{collinear} configurations. 
This type of configurations are 
defined through a single four-momentum 
$p$, and all vertex momenta 
$p_i$ $(i=1,2,3,4)$ are 
proportional to it, namely 
\be 
p_i = x_i \,p \,, 
\label{eq:configurations}
\ee
where the parameters $x_i$ are real, and  
satisfy $x_1+x_2+x_3+x_4 =0$ due to 
four-momentum conservation. 

The reason for this particular kinematic choice is twofold. First, the 
tensorial structure of the four-gluon vertex simplifies enormously, 
rendering the resulting dynamical equations completely tractable.
Second, it allows for the direct comparison with contemporary lattice simulations~\cite{Catumba:2021qbh,Colaco:2023qin,Colaco:2024gmt}, 
where these configurations 
are employed as benchmarks in the initial stages.  

Note, in fact, that,  
while the SDEs access directly the vertex 
$\fatg^{abcd}_{\mu\nu\rho\sigma}$ or 
$\overline{\fatg}^{abcd}_{\mu\nu\rho\sigma}$, 
the lattice computes the full $\mathcal{G}^{abcd}_{\mu\nu\rho\sigma}$. 
When all momenta are 
collinear, the 
disconnected contributions may be eliminated 
simply by imposing the 
additional restriction $x_i + x_j\neq 0$, 
which makes propagator-like 
transitions impossible, due to the induced 
momentum non-conservation. 
Alternatively, appropriate projectors 
may be chosen [see Sec.~\ref{sec:basis}], which 
remove the disconnected contributions from the simulation 
without restricting the values of the 
$x_i$. 
As for the 1PI terms, when 
\1eq{eq:configurations} is fulfilled, 
all three-gluon vertices in the 1PR diagrams  
are of the form 
\mbox{$\fatg_{\alpha\beta\gamma} (p)= 
A (p^2)(p_{\alpha}g_{\beta\gamma}  + p_{\beta}g_{\alpha\gamma} + p_{\gamma}g_{\alpha\beta}) + 
B (p^2)p_{\alpha} p_{\beta}p_{\gamma}$},
and therefore 
$\overline\fatg_{\alpha\beta\gamma} (p) =0$.
Thus, the term in the curly bracket of \1eq{eq:allbars}
vanishes, and 
one finally isolates  $\overline{\fatg}^{abcd}_{\mu\nu\rho\sigma}$ on the lattice~\cite{Binosi:2014kka}. 

We end this section by reporting the tree-level 
expressions of $\fatg^{abcd}_{\mu\nu\rho\sigma}$ 
and $\overline{\fatg}^{abcd}_{\mu\nu\rho\sigma}$, 
namely 
\bea
\Gamma^{abcd}_{\!0\,\mu\nu\rho\sigma} &=&  f^{adx}f^{cbx}\left(g_{\mu\rho}g_{\nu\sigma}-g_{\mu\nu}g_{\rho\sigma}\right) 
		+f^{abx}f^{dcx}\left(g_{\mu\sigma}g_{\nu\rho}-g_{\mu\rho}g_{\nu\sigma}\right) \nonumber \\ 
		&& +f^{acx}f^{dbx}\left(g_{\mu\sigma}g_{\nu\rho}-g_{\mu\nu}g_{\rho\sigma}\right) \,,
\label{Gammatree}
\eea  
where $f^{abc}$ is the totally antisymmetric SU($N$) structure constants, and 
\be  
    \overline{\Gamma}_{\!\!0~\!\!\mu\nu\rho\sigma}^{abcd}(p_{1},p_{2},p_{3},p_{4}) = 
    P^{\mu'}_\mu(p_1)P^{\nu'}_\nu(p_2)P^{\rho'}_\rho(p_3)P^{\sigma'}_\sigma(p_4) \Gamma^{abcd}_{\!0\,\mu'\!\nu'\!\rho'\!\sigma'}
    \,.
    \label{Gamma1G1}
\ee

\section{Color structure and charge conjugation symmetry}
\label{sec:color}

It is well-known that the color structure of 
the four-gluon vertex is comprised by the 
appropriate combinations of three basic 
elements, namely $f^{abc}$, the totally 
symmetric $d^{abc}$, and the identity matrix
$\delta^{ab}$. The 15 combinations built out 
of these elements are~\cite{Pascual:1980yu,Binosi:2014kka} 

\be 
f^{abx}f^{cdx}\,,\qquad d^{abx}d^{cdx}\,, \qquad f^{abx}d^{cdx} \,, \qquad \delta^{ab} \delta^{cd} \,. 
\label{Ctens}
\ee
and their permutations. The number of possibilities is reduced to 9 
thanks to 6 identities, 
valid for general SU($N$)~\cite{Pascual:1980yu,Binosi:2014kka}, 
namely 
\bea 
&& f^{abx} f^{cdx} = \frac{2}{N}[ \delta^{ac}\delta^{bd} - \delta^{ad}\delta^{bc}] + d^{acx}d^{dbx} - d^{adx} d^{bcx} \,, \nonumber\\
&& f^{abx}d^{cdx} + f^{acx}d^{dbx} + f^{adx}d^{bcx} = 0 \,, 
\label{SUN_ids}
\eea
and two permutations for each. 
Moreover, for $N = 3$, the additional identity~\cite{Pascual:1980yu}
\be 
\delta^{ab}\delta^{cd} + \delta^{ac}\delta^{bd} + \delta^{ad}\delta^{bc} = 3[d^{abx}d^{cdx} + d^{acx}d^{bdx} + d^{adx}d^{bcx}] \,, 
\label{SU3_ids}
\ee
reduces the number of independent color tensors to 8~\footnote{Note that the Jacobi identity,   \mbox{$f^{abx}f^{cdx} + f^{bcx}f^{adx} + f^{cax}f^{bdx} = 0$}, is a linear combination of the permutations of the second line in \1eq{SUN_ids}.}.

It turns out that 
the invariance of the theory under charge conjugation prohibits the presence of terms 
of the type $fd$, thus eliminating $3$ additional  color combinations.
In earlier works~\cite{Driesen:1997wz, Driesen:1998xc,Huber:2018ned}, 
and under special assumptions, 
such terms 
have been shown to vanish; however, 
no exact proof justifying their complete omission has been presented in the literature. 

The general implications of charge conjugation symmetry for the Green's functions of a pure SU($N$) theory were worked out in~\cite{Smolyakov:1980wq}. Specifically, a Green's function with $n$ color indices, $\g^{a_1 a_2\ldots a_n}$ (we suppress Lorentz indices and momenta), must satisfy
\bea 
f^{ba_1x}\g^{x a_2\ldots a_n} + f^{ba_2x}\g^{a_1 x\ldots a_n} + \ldots + f^{ba_nx}\g^{a_1 a_2\ldots x} &=& 0 \,, \nonumber\\
A^{a_1 b_1} A^{a_2 b_2} \ldots A^{a_n b_n} \g^{b_1 b_2\ldots b_n}= \g^{a_1 a_2\ldots a_n} \,. 
\label{charge}
\eea
In the above equation, $A^{ab}$ is a certain color matrix that conjugates the charges; its concrete form will not be necessary here. It suffices to state the following properties of $A^{ab}$:
\begin{enumerate}[label={\{\arabic*\}}]
\item $A^{ab} \not\propto \delta^{ab}$, \ie $A^{ab}$ is not proportional to the diagonal matrix. 
\item $A$ is an orthogonal matrix, \ie $A^\intercal = A^{-1}$, where $\intercal$ denotes matrix transposition; \label{prop1}
\item $A^{aa^\prime}A^{bb^\prime}A^{cc^\prime} f^{a^\prime b^\prime c^\prime} = f^{abc}$; \label{prop2}
\item $A^{aa^\prime}A^{bb^\prime}A^{cc^\prime} d^{a^\prime b^\prime c^\prime} = -d^{abc}$. \label{prop3}
\end{enumerate}

To see how \1eq{charge} can then be used to constrain the color structures of vertex functions, consider first the simpler case of the three-gluon vertex, $\fatg_{\alpha\mu\nu}^{abc}(q,r,p)$. The most general structure possible is given by~\cite{Pascual:1980yu}
\be 
\fatg_{\alpha\mu\nu}^{abc}(q,r,p) = f^{abc} V_{\alpha\mu\nu}(q,r,p) + d^{abc} U_{\alpha\mu\nu}(q,r,p) \,, 
\label{3g_general}
\ee
where the tensors 
$V_{\alpha\mu\nu}(q,r,p)$ and $U_{\alpha\mu\nu}(q,r,p)$ may be decomposed 
in an appropriate basis.

Using the Jacobi identity  and the second line of \1eq{SUN_ids}, 
it is easy to verify   
that the expression in 
\1eq{3g_general}  satisfies the first line of \1eq{charge}.
Then, substituting \1eq{3g_general} into the second line of \1eq{charge} and using properties \ref{prop2} and \ref{prop3}, we obtain
\be 
d^{abc} U_{\alpha\mu\nu}(q,r,p) = 0 \qquad \Longrightarrow \qquad U_{\alpha\mu\nu}(q,r,p) = 0\,.
\ee
Hence, charge conjugation symmetry implies that the three-gluon vertex is proportional to $f^{abc}$ only.

Let us now consider the four-gluon vertex. From property \ref{prop1} follows that 
\be 
A^{xa}A^{xb} = A^{ax}A^{bx} = \delta^{ab} \,,
\ee
and thus  
\be 
f^{abx}f^{cdx} = f^{abx}\delta^{xy}f^{cdy} = f^{abx}(A^{zx}A^{zy})f^{cdy} \,.
\ee
Hence, multiplying by $A^{a a^\prime}A^{b b^\prime}A^{c c^\prime}A^{d d^\prime}$, we find
\bea 
A^{a a^\prime}A^{b b^\prime}A^{c c^\prime}A^{d d^\prime} f^{abx}f^{cdx} &=& A^{a a^\prime}A^{b b^\prime}A^{c c^\prime}A^{d d^\prime} f^{abx}(A^{zx}A^{zy})f^{cdy} \nonumber\\
&=& (A^{a a^\prime}A^{b b^\prime}A^{zx} f^{a^\prime b^\prime x})( A^{c c^\prime}A^{d d^\prime}A^{zy}f^{c^\prime d^\prime y})= f^{abz}f^{cdz} \,, \label{transf_ff}
\eea
where we used property \ref{prop2} to obtain the last equality. 

Similarly, using property \ref{prop1}, we get 
\be 
A^{a a^\prime}A^{b b^\prime}A^{c c^\prime}A^{d d^\prime} \delta^{a^\prime b^\prime} \delta^{c^\prime d^\prime} = \delta^{ab}\delta^{cd} \,. \label{transf_deltadelta}
\ee

Finally, using property \ref{prop3}, it is 
straightforward to show that
\bea 
A^{a a^\prime}A^{b b^\prime}A^{c c^\prime}A^{d d^\prime} d^{a^\prime b^\prime e} d^{c^\prime d^\prime e} &=& d^{abe} d^{cde}  \,, \nonumber\\
A^{a a^\prime}A^{b b^\prime}A^{c c^\prime}A^{d d^\prime} f^{a^\prime b^\prime e} d^{c^\prime d^\prime e} &=& - f^{abe} d^{cde} \,. \label{transf_dd_fd}
\eea

Combining the transformation properties given by \3eqs{transf_ff}{transf_deltadelta}{transf_dd_fd} into the second line of \1eq{charge}, one can easily show that the form factors of the color structures of the form $fd$ must all vanish.

Consequently, the number of independent color structures in the four-gluon vertex is reduced to $5$, namely  
\bea 
f^{abx}f^{cdx}\,;  \qquad  f^{acx}f^{bdx}\,; \qquad 
\delta^{ab} \delta^{cd} \,;  \qquad 
\delta^{ac} \delta^{bd} \,;  \qquad \delta^{ad} \delta^{bc}\,.  
\label{colorTens}
\eea

\section{Tensor basis and projectors}
\label{sec:basis}

In this section we discuss the tensorial basis 
that will be used for describing the 
four-gluon vertex in collinear configurations, 
and the projectors that allow us to extract 
particular form factors. 

From this point on, we will employ 
the short-hand notation
\be
A(x,p) := A(x_1p,x_2p,x_3p,x_4p)  \,,
\ee
to indicate a quantity $A$  
(with Lorentz and color indices, as well as 
scalar form factors) 
in general collinear kinematics. 
In the case of a specific configuration,
\ie $(1,1,1,-3)$,  we will be reverting to the 
explicit notation, \ie $A(p,p,p,-3p)$.

First, notice that in collinear configuration one can form 10 Lorentz tensors with four indices and one independent momentum~\cite{Cyrol:2014kca,Gracey:2014ola}, covering all linearly independent combinations of the forms
\bea 
g^{\mu\nu}g^{\rho\sigma}\;\mbox{(3 tensors)}\,; \qquad  
g^{\mu\nu}p^{\rho}p^{\sigma}\;\mbox{(6 tensors)}\,;  \qquad
p^{\mu}p^{\nu}p^{\rho}p^{\sigma} \;\mbox{(1 tensor)}\,. 
\label{Ltens}
\eea

Since $P_{\mu\nu}(x_i p) = P_{\mu\nu}(p)$, for any scalar $x_i$, the Lorentz tensors quadratic and quartic in the momenta [cf.~\1eq{Ltens}] do not survive the transverse projection~\eqref{4g_transv} in the collinear configurations defined in Eq.~\eqref{eq:configurations}. Hence, the Lorentz tensors available to decompose $\overline{\fatg}^{abcd}_{\mu\nu\rho\sigma}(p_1, p_2, p_3, p_4)$ reduce to only 3, namely
\be 
P_{\mu\nu}(p)P_{\rho\sigma}(p) \,, \qquad P_{\mu\rho}(p)P_{\nu\sigma}(p)\,, \qquad P_{\mu\sigma}(p)P_{\nu\rho}(p) \,.
\label{LorTens}
\ee

Combining the 5 independent color tensors of \1eq{colorTens} with the Lorentz tensors of \1eq{LorTens}, we see that for collinear configurations $\overline{\fatg}^{abcd}_{\mu\nu\rho\sigma}$ is comprised by 15 color-Lorentz tensors. However, the basis resulting from the tensor product of the building blocks in \2eqs{colorTens}{LorTens} is not manifestly Bose symmetric.

To amend this, we 
use linear combinations of the above elements to 
construct a basis of 15 new tensors $t_{i,\mu\nu\rho\sigma}^{abcd}$, and associated form factors $G_i$, 
\be 
\overline{\fatg}^{abcd}_{\mu\nu\rho\sigma}(x,p) = \sum_{i = 1}^{15} \T_i (x,p)  \, t_{i,\mu\nu\rho\sigma}^{abcd} (x,p)\,. 
\label{basis}
\ee

Notice that in the decomposition given by Eq.~\eqref{basis}, one has that
\begin{enumerate}[label=({\it \roman*})]
\item Each $t_i$ 
is manifestly Bose symmetric. Consequently, the form factors $\T_i$ are individually symmetric under the exchange of any two components $x_i$ implicit in $x$.

\item The $t_i$, and therefore the $\T_i$, are all dimensionless. This property prevents the form factors $\T_i$ from developing kinematic divergences when one of the momenta vanishes. Moreover, the different form factors can be compared directly.

\item Finally, the tree-level tensor $\gbar^{0}$
defined in \1eq{Gamma1G1} is an element of the basis, namely 
$t_1 = 
\gbar^{0}$.
\end{enumerate}

To construct a basis satisfying the above properties, we employ the $S_4$ permutation group formalism developed in~\cite{Eichmann:2015nra}. Specifically, we combine the multiplets of momenta, colors, and Lorentz indices computed there into $S_4$ singlets (invariants). The details are presented in Appendix~\ref{basis_col}, and 
the expressions for the resulting $t_i$ are given in~\1eq{tibasis}.

In order to extract the form factors $\T_k$ from the SDE, we construct projectors $\mathcal{P}_k$ that isolate specific elements of the basis. In compact notation, we have
\begin{align}
    \mathcal{P}_{k} \odot \overline{\fatg} = \T_{k}\,, ~~~~~k = 1,\ldots, 15 \,,
    \label{Projdef}
\end{align}
where the symbol  ``$\odot$''
denotes the full contraction 
of all tensor and color 
indices, namely \mbox{$A \odot B 
= A^{abcd}_{\mu\nu\rho\sigma} 
B_{abcd}^{\mu\nu\rho\sigma}$}.

To determine the projectors $\mathcal{P}_{k}$ , we first contract both sides of \1eq{basis} with an arbitrary $t_{j}$,
\be 
t_{j}\odot\overline{\fatg} = \sum_{i = 1}^{15} \T_i~ (t_{j}\odot t_{i}) = \sum_{i = 1}^{15} \T_i~ [C]_{ji} \,, ~~~~~j = 1,\ldots, 15  \,, \label{eq:proj_step1}
\ee
where we define the symmetric matrix $C$ with entries $[C]_{ji} = (t_{j}\odot t_{i})$. Next we multiply both sides of \1eq{eq:proj_step1} by the inverse $[C^{-1}]_{kj}$ thus obtaining
\be
\sum_{j = 1}^{15}([C^{-1}]_{kj} ~t_{j})\odot \overline{\fatg} = \T_{k} \implies \mathcal{P}^{abcd}_{k,\mu\nu\rho\sigma} : = \sum_{j = 1}^{15}[C^{-1}]_{kj} ~t^{abcd}_{j,\mu\nu\rho\sigma}\,, 
\label{eq:proj}
\ee
with $k = 1,\ldots, 15$.

The first three tensors, $t_{1,2,3}$, of the basis of \1eq{basis} form a rather special subset. To begin, they are orthogonal to the remaining tensors, \ie
\begin{align}
    t_{1}\odot t_{j} = t_{2}\odot t_{j} = t_{3}\odot t_{j} = 0\,,~~~~~~j=4,\ldots,15\,.
    \label{eq:orthog}
\end{align}
In fact, $t_{3}$, is orthogonal to all other $t_{i}$. Furthermore, we find that $t_{1,2,3}$ are the only tensors of the basis that are independent of $x_{i}$\footnote{The tensors $t_{14}$ and $t_{15}$ are also orthogonal to all other tensors of the basis, but they depend on the  $x_{i}$.}. Indeed, for $t_{1,2,3}$ the expressions in~\1eq{tibasis} reduce to the relatively compact forms
\begin{align}
     t_{1\,\mu\nu\rho\sigma}^{abcd} & = \gbar^{0\,abcd}_{\mu\nu\rho\sigma} \,,\nonumber\\
     t_{2\,\mu\nu\rho\sigma}^{abcd} &=\delta^{ad} \delta^{bc} \left[ 2 P_{\mu\sigma}(p)P_{\nu\rho}(p)
    -P_{\mu\rho}(p)P_{\nu\sigma}(p)
    -P_{\mu\nu}(p) P_{\rho\sigma}(p) \right]
     \nonumber\\
     &\,+\delta^{ac} \delta^{bd}\left[ 2 P_{\mu\rho}(p)P_{\nu\sigma}(p)
    -P_{\mu\sigma}(p)P_{\nu\rho}(p)
    -P_{\mu\nu}(p) P_{\rho\sigma}(p) \right] \nonumber\\
     &\,+\delta^{ab}\delta^{cd} \left[ 2 P_{\mu\nu}(p)P_{\rho\sigma}(p)
    -P_{\mu\rho}(p)P_{\nu\sigma}(p)
    -P_{\mu\sigma}(p) P_{\nu\rho}(p) \right]\,, \nonumber\\
    t_{3\,\mu\nu\rho\sigma}^{abcd} &= ( \delta^{ab}\delta^{cd} + \delta^{ac}\delta^{bd} + \delta^{ad}\delta^{bc} )\left[ P_{\mu\nu}(p) P_{\rho\sigma}(p) + P_{\mu\rho}(p)P_{\nu\sigma}(p) + P_{\mu\sigma}(p)P_{\nu\rho}(p) \right]\,. 
\label{basis_ti}    
\end{align}

These two observations have important consequences for the projectors $\mathcal{P}_{1}, \mathcal{P}_{2}$ and $\mathcal{P}_{3}$ which extract the associated form factors. The orthogonality expressed in \1eq{eq:orthog} gives rise to 
a significant simplification in the procedure of \1eq{eq:proj}, allowing these projectors to be expressed as linear combinations of only $t_{1,2,3}$; in particular, $\mathcal{P}_{3}$ is simply proportional to $t_{3}$. As a consequence, we find that $\mathcal{P}_{1}, \mathcal{P}_{2}$, and $\mathcal{P}_{3}$, are themselves independent of $x_{1}, x_{2}$ and $x_{3}$; through the procedure outlined, we find that they read 
\be
\mathcal{P}_1 = \displaystyle \frac{ 7 t_1 - 3 t_2}{19440}\,;  \qquad  \mathcal{P}_2  
= \frac{  3 t_2  - 2 t_1 }{12960} \,; \qquad 
\mathcal{P}_3  = \frac{ t_3}{10800} \,.
\label{projs123}
\ee
We hasten to emphasize that, since we made explicit use of \1eq{SU3_ids}, the above projectors are valid for SU(3), and may not hold for a general SU($N$) group.

In what follows we will concentrate 
on the form factors $\T_{1,2,3}(x,p)$, 
since they do not mix with the remaining form factors in any collinear configuration. 

 We conclude by pointing out that the projector $\mathcal{P}_1$ completely eliminates disconnected contributions when applied to $\mathcal{G}^{abcd}_{\mu\nu\rho\sigma}$, because its contraction with color structures of the type $\delta^{ab}\delta^{cd}$ vanish. Thus, in principle, the tree-level tensor structure can be isolated on the lattice, even for $x_i + x_j = 0$, by employing  $\mathcal{P}_1$. On the other hand, the tensors $t_{2,3}$ mix with disconnected terms in $\mathcal{G}^{abcd}_{\mu\nu\rho\sigma}$, since the projectors $\mathcal{P}_{2,3}$ do not annihilate $\delta^{ab}\delta^{cd}$. 

\section{SDE of the four-gluon vertex}
\label{sec:sde}

In this section, we set up the  SDE that
governs the form factors $\T_{1,2,3}$  of the four-gluon vertex for an arbitrary collinear configuration, and discuss 
the renormalization of the resulting equations. 

\begin{figure}[ht]
 \includegraphics[width=0.95\linewidth]{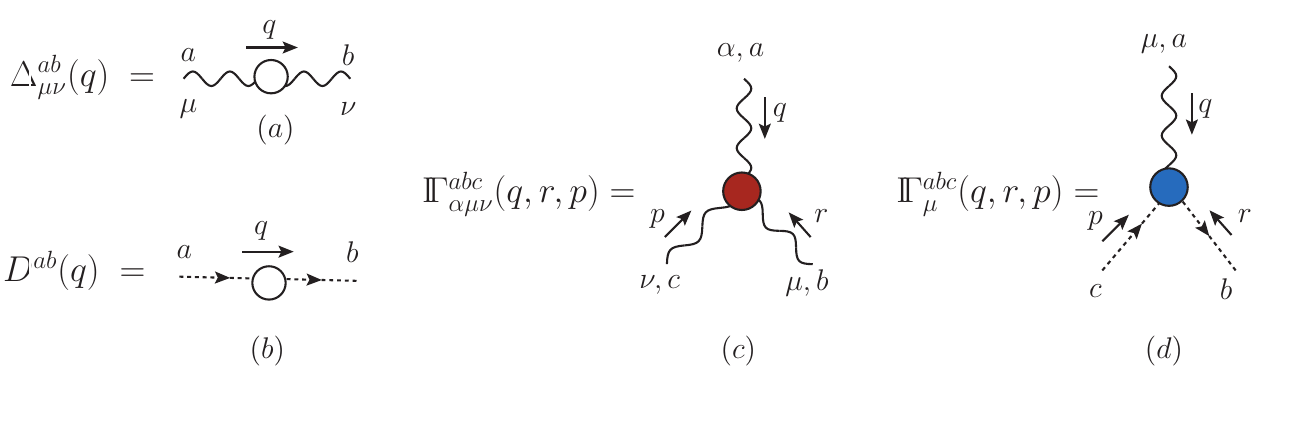}
 \caption{Diagrammatic representations of: ($a$) the fully dressed gluon propagator,  \mbox{$\Delta_{\mu\nu}(q)$};  ($b$) the complete ghost propagator, \mbox{$D(q^2)$}; ($c$) the full three-gluon vertex, \mbox{$\fatg^{abc}_{\alpha\mu\nu}(q,r,p) =g f^{abc}\fatg_{\alpha\mu\nu}(q,r,p)$}; ($d$) the full ghost-gluon vertex,  $\fatg^{abc}_\mu(q,r,p) =-g f^{abc} \fatg_\mu(q,r,p)$.} 
\label{fig:prop_vert}
\end{figure}

We employ the version of the four-gluon SDE derived from the formalism of the 4PI effective action~\cite{Cornwall:1974vz,Cornwall:1973ts,Berges:2004pu,York:2012ib,Williams:2015cvx} at the {\it four-loop} level~\cite{Carrington:2010qq,Carrington:2013koa}; 
we remind the reader that, 
within this formalism,  
the SDE of a given 
Green's function 
is obtained 
by extremizing the variation 
of the effective action with respect to this 
particular 
function.

\begin{figure}[t]
 \includegraphics[width=0.95\linewidth]{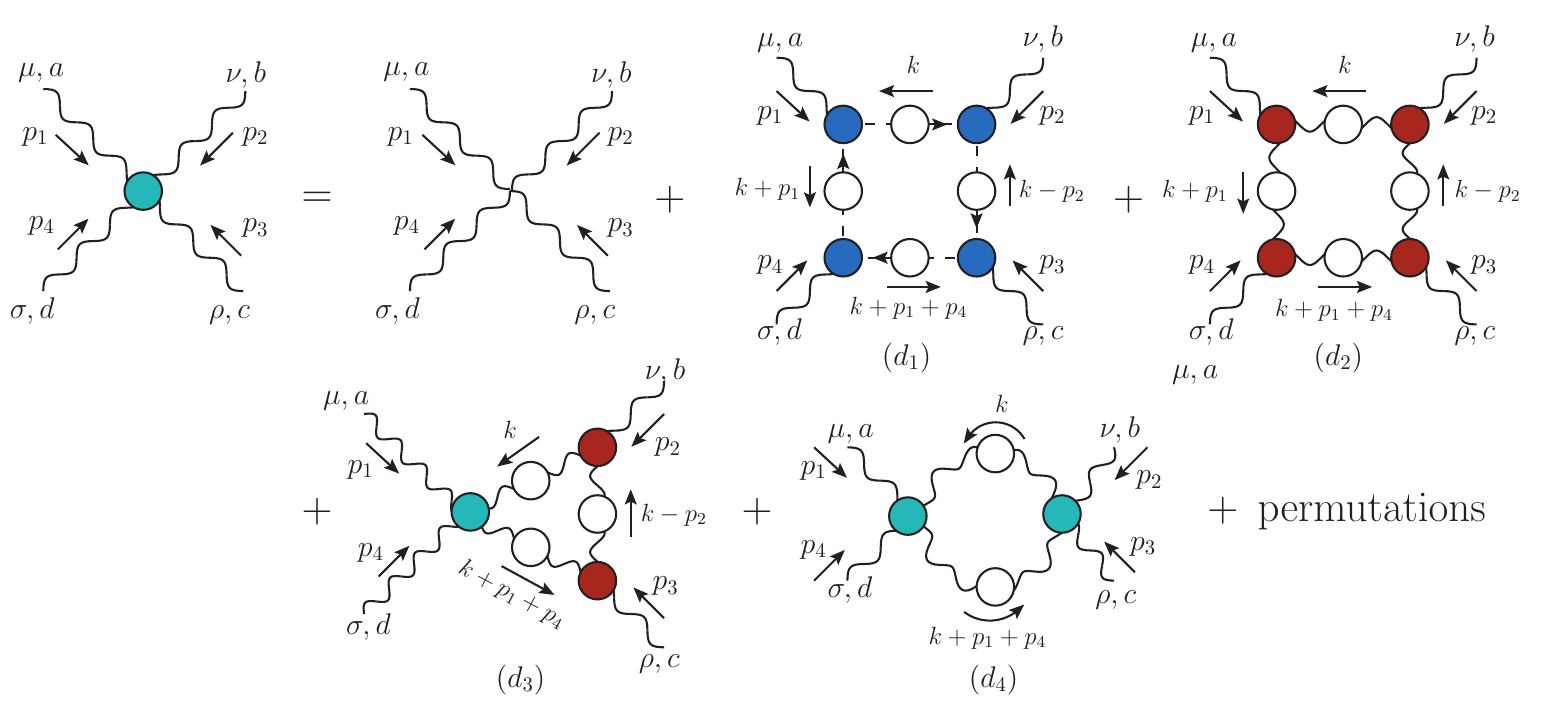}
 \caption{Diagrammatic representation of the SDE for the full four-gluon vertex, ${\fatg}^{abcd}_{\mu\nu\rho\sigma}$, derived from the 
 4PI effective action at the four-loop level. Contributions obtained through the permutations of the external momenta in the various diagrams are omitted.}
\label{fig:SDE4g}
\end{figure}

 It turns out that the SDE of the four-gluon vertex 
derived from the 4PI effective action 
is automatically symmetric with respect to permutations of its external legs, and contains 
fully dressed propagators and vertices as its 
ingredients, depicted diagrammatically  
in Fig.~\ref{fig:prop_vert}.
 The four-gluon SDE built out of these 
 components is shown in Fig.~\ref{fig:SDE4g}; 
 diagram $(d_1)$ is accompanied by five   additional permutations,  
 $(d_2)$ by two, $(d_3)$ by five, and 
 $(d_4)$ by two~\cite{Binosi:2014kka}. 
 In what follows we will denote by 
\begin{align}
\left({d}_{i}^{s}\right)^{abcd}_{\mu\nu\rho\sigma}(p_{1},p_{2},p_{3},p_{4})= \left({d}_{i}\right)^{abcd}_{\mu\nu\rho\sigma}(p_{1},p_{2},p_{3},p_{4}) +  \mbox{permutations}\,,
 \label{eq:dsum}
 \end{align}
the sum of 
each representative diagram and its permutations.

To obtain the SDE of the 
transversely-projected four-gluon vertex, $\overline{\fatg}^{abcd}_{\mu\nu\rho\sigma}(p_1,p_2,p_3,p_4)$, one simply 
contracts both sides of the SDE 
in Fig.~\ref{fig:SDE4g} by the transverse 
projectors corresponding to the four external legs;  the additional projectors,  needed for the internal lines, come automatically from the gluon propagators, which in the Landau gauge, assumes the completely transverse form, $\Delta_{\mu\nu}(q) = \Delta(q^2)P_{\mu\nu}(q)$.
Thus, in addition to $\overline{\fatg}^{abcd}_{\mu\nu\rho\sigma}$, 
the various diagrams depend on the transversely-projected three-gluon
and ghost-gluon vertices,
$\overline{\fatg}^{\,\alpha\mu\nu}(q,r,p)$ 
and 
$\overline{\fatg}^{\,\mu}(q,r,p)$, 
respectively, defined as 
\be
\overline{\fatg}^{\,\alpha\mu\nu}(q,r,p) := P_{\alpha'}^{\alpha}(q) P_{\mu'}^{\mu}(r) P_{\nu'}^{\nu}(p) \fatg^{\alpha'\!\mu'\!\nu'}(q,r,p) \,, \,\,\,\,\,\,
\overline{\fatg}^{\,\mu}(q,r,p) := P_{\mu'}^{\mu}(q) \fatg^{\mu'}(q,r,p) \,.
\ee
Then, the transversely-projected diagrams $(d_i)^{abcd}_{\mu\nu\rho\sigma}$ 
in Fig.~\ref{fig:SDE4g}, to be denoted by $(\bar{d}_{i})^{abcd}_{\mu\nu\rho\sigma}$,
are given by (Minkowski space, omitting a factor 
$-ig^{2}$) 

 \begin{align}
     \label{dis4g} 
    \left(
 \bar{d}_{1}\right)^{abcd}_{\mu\nu\rho\sigma} &=  \!\! -2\!\!\int_k \!\!
     \overline{\fatg}^{amf}_{\mu}\!(p_{1},k,-k_{1})  
     \overline{\fatg}^{bnm}_{\nu}\!(p_{2},k_{2},-k) 
     \overline{\fatg}^{cnt}_{\rho}\!(p_{3},-k_{2},k_{3})
     \overline{\fatg}^{dft}_{\sigma}\!(p_{4},k_{1},-k_{3})f_{D}(k,k_{1},k_{2},k_3)\,, \nonumber\\
 \left(\bar{d}_{2}\right)^{abcd}_{\mu\nu\rho\sigma} &=   \!\!\int_k \!\!
     \overline{\fatg}_{\mu\alpha\tau}^{amf}(p_{1},k,-k_{1}) 
     \overline{\fatg}_{\nu\beta\alpha}^{bnm}(p_{2},k_{2},-k) 
     \overline{\fatg}_{\rho\beta\lambda}^{cnt}(p_{3},-k_{2},k_{3}) 
     \overline{\fatg}_{\sigma\tau\lambda}^{dft}(p_{4},k_{1},-k_{3})f_{\Delta}(k,k_{1},k_{2},k_3)\,,  \nonumber \\
 \left(\bar{d}_{3}\right)^{abcd}_{\mu\nu\rho\sigma} &=  \!\!\int_k 
     \overline{\fatg}_{\nu\tau\alpha}^{btm}(p_{2},k_{2},-k)    
     \overline{\fatg}_{\rho\beta\tau}^{cnt}(p_{3},k_{3},-k_{2})  \overline{\fatg}^{admn}_{\mu\sigma\alpha\beta}(p_{1},p_{4}, k, -k_{3}) 
 \Delta(k^2)\Delta(k_{3}^2)\Delta(k_{2}^2)\,,\nonumber\\
 \left(\bar{d}_{4}\right)^{abcd}_{\mu\nu\rho\sigma} &=  \!\! \frac{1}{2}\int_k  
 \overline{\fatg}^{admn}_{\mu\sigma\alpha\beta}(p_{1},p_{4},k,-k_{3}) 
 \overline{\fatg}^{bcmn}_{\nu\rho\alpha\beta}(p_{2},p_{3},-k,k_{3}) 
 	\Delta(k^2) \Delta(k_{3}^2) \,,
 \end{align} 
where we set $k_1:=k+p_{1}$; $k_{2}:=k-p_{2}$;    and $k_{3}:=k+p_{1}+p_{4}$.  In addition, we 
define the string of propagators 
\bea
f_{{\mathcal D}}(k,k_{1},k_{2},k_3) &:=&{\mathcal D}(k^2){\mathcal D}(k_{1}^2){\mathcal D}(k_{2}^2){\mathcal D}(k_{3}^2) \,, \,\,\,\,\,\,  {\rm with} \,\,\, {\mathcal D}(q^2) 
= \Delta(q^2), D(q^2) \,,
\eea
where \mbox{ $D^{ab}(q^2)= i\delta^{ab}D(q^2)$} is 
the ghost propagator, whose dressing 
function  is given by \mbox{$F(q^2) = q^2 D(q^2)$}.

Finally, 
\be
\label{eq:int_measure}
\int_{k} := \frac{1}{(2\pi)^4} \int \!\!{\rm d}^4 k \,,
\ee
where the use of a symmetry-preserving regularization scheme is implicitly assumed.

Thus, the SDE for $\overline{\fatg}^{abcd}_{\mu\nu\rho\sigma}$
may be expressed in the following compact way
\begin{align}
\label{4gsde}
\overline{\fatg}^{abcd}_{\mu\nu\rho\sigma}(p_{1},p_{2},p_{3},p_{4}) = \overline{\Gamma}_{\!0\,\mu\nu\rho\sigma}^{abcd}(p_{1},p_{2},p_{3},p_{4}) \,
+ \sum_{i=1}^{4} \left(
\bar{d}_{i}^s\right)^{abcd}_{\mu\nu\rho\sigma}(p_{1},p_{2},p_{3},p_{4}) \,,
\end{align} 
where the $(\bar{d}_{i}^s)$ denotes 
the transversely-projected counterpart of the 
$(d_i^s)$ defined in \1eq{eq:dsum}.

We now turn to 
the renormalization of the SDE given in  
\1eq{dis4g}.  
All quantities appearing in that equation 
are bare (unrenormalized);  
the transition to the renormalized quantities is implemented multiplicatively, 
by means of the standard relations 
\begin{align} 
\Delta_{\s R}(q^2)&= Z^{-1}_{A} \Delta(q^2), &\quad\quad\quad \fatg^{\mu}_{\!\!\s R}(q,p,r) &= Z_1 \fatg^{\mu}(q,p,r),\nonumber\\  
F_{\!\s R}(q^2)&= Z^{-1}_{c} F(q^2),&\quad\quad\quad \fatg^{\alpha\mu\nu}_{\!\!\s R}(q,r,p) &=  Z_3 \fatg^{\alpha\mu\nu}(q,r,p),\nonumber\\ 
g_{\s R} &= Z_g^{-1} g\,, &\quad\quad\quad  \fatg^{abcd}_{\!\!\s R\,\alpha\beta\mu\nu}(q,r,p,t) &=  Z_4 \fatg^{abcd}_{\alpha\beta\mu\nu}(q,r,p,t)\,,
\label{renconst}
\end{align} 
where the subscript ``$R$'' denotes renormalized quantities, and  $Z_{A}$, $Z_{c}$, $Z_{1}$, $Z_{3}$, $Z_{4}$, and $Z_g$ are the corresponding (cutoff-dependent) renormalization constants.
In addition, we employ the exact relations 
\be
Z_g^{-1} = Z_1^{-1} Z_A^{1/2} Z_c\, = Z_3^{-1} Z_A^{3/2} \, = Z_4^{-1/2} Z_A \,,
\label{eq:sti_renorm}
\ee 
which are imposed by the fundamental Slavnov-Taylor identities (STIs)~\cite{Taylor:1971ff,Slavnov:1972fg}. 
Substituting the relations of \1eq{renconst} into \1eq{dis4g}
and using \1eq{eq:sti_renorm}, it is straightforward to 
 derive  the renormalized version of \1eq{4gsde}, given by  
\begin{align}
\label{renorm_4gsde}
\overline{\fatg}^{abcd}_{{\!\!\s R}\, \mu\nu\rho\sigma}(p_{1},p_{2},p_{3},p_{4}) = Z_4\overline{\Gamma}_{\!0\,\mu\nu\rho\sigma}^{abcd}(p_{1},p_{2},p_{3},p_{4}) \,
+  \sum_{i=1}^{4} \left(
\bar{d}^{s}_{i\,\s R}\right)^{abcd}_{\mu\nu\rho\sigma} (p_{1},p_{2},p_{3},p_{4})\,,
\end{align}
where the 
subscript ``$R$'' in $ \left(
\bar{d}^{s}_{i\,\s R}\right)^{abcd}_{\mu\nu\rho\sigma}$ 
denotes that the expressions given in 
\1eq{dis4g} have been substituted by their renormalized counterparts. We emphasize that the $Z_4$ multiplied by the tree-level structure is the only renormalization constant to be determined. 
Note in particular that, because all vertices are fully dressed,  none of the renormalization 
constants appears multiplying  
any of the contributions 
$ \left(
\bar{d}^{s}_{i\,\s R}\right)^{abcd}_{\mu\nu\rho\sigma}$; thus, the usual complications 
associated with multiplicative 
renormalization are absent~\cite{Carrington:2010qq,Carrington:2013koa}.

In order to determine $Z_4$ we adopt 
a variant of the momentum subtraction (MOM) scheme~\cite{Boucaud:2008gn,Boucaud:2011eh}.
Within this scheme, the gluon and ghost propagators assume their tree-level values at the subtraction point $\mu$, \ie 
\be 
\Delta^{-1}(\mu^2) = \mu^2 \,, \qquad F(\mu^2) = 1 \,.  
\label{MOM}
\ee
In general studies of vertices, these two conditions 
are supplemented by an additional one, 
fixing the value of a form factor (or combination thereof)  
at some special kinematic configuration~\cite{Aguilar:2020yni,Aguilar:2020uqw,Aguilar:2021lke,Aguilar:2020uqw}. 
For example, in the recent SDE 
analysis of~\cite{Aguilar:2023qqd}, the renormalization 
of the three-gluon vertex was such that the 
leading form factor would acquire its tree-level 
value in the soft-gluon configuration [see Eq.~(3.13) therein].

For the case of the four-gluon vertex we employ an analogous 
scheme. In particular, 
we single out the configuration 
$(0,p,p,-2p)$, and require that 
\be 
\qquad \T_{1}(0,p,p,-2p)\rvert_{p^{2}=\mu^{2}}  = 1 \,.  
\label{ren_conds_4g}
\ee
There is nothing special about this 
configuration, other than its 
relative numerical simplicity; any other choice would be  
equally good for renormalizing the 
SDE.

Note that, when 
the scheme defined by 
\2eqs{MOM}{ren_conds_4g} is employed, the conditions used 
for renormalizing the three-gluon and ghost-gluon vertices may {\it not} be simultaneously employed, because this would lead to 
a violation of the relations of 
\1eq{eq:sti_renorm}
imposed by the STIs~\cite{Celmaster:1979km}. 
The renormalized three-gluon and ghost-gluon vertices 
used in our calculations as inputs of the four-gluon SDE 
must undergo finite renormalizations, 
which will adjust their values such that the relations
of \1eq{eq:sti_renorm}
are fulfilled. This procedure 
of finite renormalization 
is described in detail in  
Appendix~\ref{scheme}.


The determination of $Z_{4}$ 
proceeds by first employing 
the special projection of \1eq{projs123} to define  
\be 
{d}^{\,0}_{i}(p^{2}) :=\left[\displaystyle  \mathcal{P}_{1} \odot (\bar{d}_{i\s R}^{s})(x,p) \right]\Big|_{x \to x_0}\,;
\label{eq:Ldiagram}
\ee
where $x_0$ denotes 
the reference configuration $(0,1,1,-2)$ 
in \1eq{ren_conds_4g}. Then, imposing the condition 
\1eq{ren_conds_4g} at the level of \1eq{renorm_4gsde},  we find
\be
Z_{4} = 1 -  \sum_{i=2}^{4} {d}^{\,0}_{i}(\mu^{2}) \,.
\label{Z4equation}
\ee

Note that all contributions to ${d}^{\,{0}}_{1}(p^{2})$ originating from the ghost boxes \emph{vanish}, \ie 
\mbox{$\left[\displaystyle  \mathcal{P}_{1} \odot (\bar{d}_{1\s R}^{s})(x,p) \right]\Big|_{x \to x_0} \!\!\!= 0 $}.

Then, substituting Eq.~\eqref{Z4equation} into Eq.~\eqref{renorm_4gsde}, we obtain the renormalized SDE for $\overline{\fatg}^{abcd}_{{\!\!\s R}\, \mu\nu\rho\sigma}$ 
expressed as
\begin{align}
\label{renorm_4gsdef}
\overline{\fatg}^{abcd}_{{\!\!\s R}\, \mu\nu\rho\sigma}(x,p) = \overline{\Gamma}_{\!0\,\mu\nu\rho\sigma}^{abcd}(x,p) \left[
1 - \sum_{i=2}^{4} {d}^{\,0}_{i}(\mu^{2})
\right] \,
+  \sum_{i=1}^{4} \left(
\bar{d}^{s}_{i\,\s R}\right)^{abcd}_{\mu\nu\rho\sigma} (x,p)\,.
\end{align}
When no ambiguity can arise we drop the index “R” to avoid notational clutter. 

The final 
renormalized expressions for both $\T_{1,3}(x,p)$ 
are obtained by acting  
on \1eq{renorm_4gsdef}
with the associated projectors given in \1eq{projs123}, 
\bea
\T_{1}(x,p) &=& 
1 
+  \sum_{i=2}^{4} \mathcal{P}_{1}\odot \left(\bar{d}_{i}^{s}\right)(x,p)-  \sum_{i=2}^{4} {d}^{\,{0}}_{i}(\mu^{2})\,,\nonumber\\
\T_{j}(x,p) &=&  
 \sum_{i=1}^{4} \mathcal{P}_{j}\odot \left(\bar{d}_{i}^{s}\right)(x,p)\,,   \qquad j=2,3\,.
\label{G123eqs}
\eea

The SDEs in \1eq{G123eqs} 
will be solved in the next section, 
under a set of well-founded 
assumptions, which will be 
discussed there in detail.
It turns out that, 
under these assumptions,  
one may demonstrate analytically 
the ultraviolet finiteness of the 
resulting $G_i(x,p)$; the detailed 
proof is presented in 
Sec.~\ref{sec:Gi_finite}.

\section{Results: Emergence of Planar degeneracy}
\label{sec:num}

In this section we solve the SDE of the four-gluon vertex 
under certain simplifying assumptions, and 
check the extent of validity of the planar degeneracy 
at the level of the solutions obtained. 

\subsection{Preliminary considerations}\label{sec:numA}

For the SDE treatment, 
we employ appropriate inputs for the various functions entering in the kernels of \2eqs{eqG1}{eqG3}. In particular, 
for the gluon propagator, $\Delta(r^2)$, 
and the ghost dressing function, $F(r^2)$, 
we use fits given by  Eqs.~(C11) and (C6) of~\cite{Aguilar:2021uwa}, respectively, to the lattice results of~\cite{Bogolubsky:2009dc,Aguilar:2021okw}.

Regarding the three-gluon vertex, 
we retain only its tree-level tensorial structure, and resort to 
the planar degeneracy approximation 
for the associated form factor. 
Specifically, 
$\overline{\fatg}^{\,\alpha\mu\nu}(q,r,p)$ can be accurately approximated by the compact form
\be 
\overline{\fatg}^{\,\alpha \mu \nu}(q,r,p) =\Ls(s^2) \overline{\g}_{\!0}^{\,\alpha \mu \nu}(q,r,p) \,, \qquad s^2= \frac{1}{2}(q^2+r^2+p^2)\,,
\label{compact}
\ee
where $\overline{\g}_{\!0}^{\,\alpha \mu \nu}(q,r,p) = \ P_{\alpha'}^\alpha(q)  P_{\mu'}^{\mu}(r)  P_{\nu'}^\nu(p) 
\g_{\!0}^{\,\alpha' \mu' \nu'}(q,r,p)$,  with $\g_{\!0}$ denoting the three-gluon vertex at tree-level~\cite{
Eichmann:2014xya,Ferreira:2023fva,
Pinto-Gomez:2022brg,Aguilar:2023qqd}. 
The function $\Ls(s^{2})$ is 
the form factor associated with the soft-gluon limit of the three-gluon
vertex, $(q=0, r =-p)$, and has been accurately determined from various lattice simulations~\mbox{\cite{Athenodorou:2016oyh,Duarte:2016ieu,Boucaud:2017obn,Pinto-Gomez:2022brg,Aguilar:2019uob,Aguilar:2021lke,Aguilar:2021okw}}. Concretely, we use the fit given by Eq.~(C12) of~\cite{Aguilar:2021uwa}, to the lattice data of~\cite{Aguilar:2022thg,Ferreira:2023fva}, and convert to the renormalization scheme of \2eqs{MOM}{ren_conds_4g} using \1eq{scheme_conversion_fin}.

As for the ghost-gluon vertex, 
its Lorentz decomposition reads
\be 
\fatg_\mu(q,r,p) = r_\mu B_1(q,r,p) + q_\mu B_2(q,r,p) \,;
\ee
at tree-level, $B_1^0 = 1$ and $B_2^0 = 0$. Only $B_1$ contributes to the transversely-projected form, $\overline{\fatg}^{\,\mu}(q,r,p)$, so we have 
\begin{align}
\overline{\fatg}^{\,\mu}(q,r,p) &:= r^{\mu'}P_{\mu'}^{\mu}(q) B_1(q,r,p) \,.
\end{align}
The form employed for 
$B_{1}(q,r,p)$ is taken from the SDE results of~\cite{Aguilar:2022thg,Ferreira:2023fva} (see Fig.~13 of \cite{Ferreira:2023fva}), and has been consistently renormalized using \1eq{scheme_conversion_fin}.

Finally, for the coupling constant we use the value \mbox{$\alpha_s(\mu^2) = g^2/4\pi = 0.23$} [see \1eq{scheme_conversion_fin}], 
which corresponds to the renormalization scheme of \2eqs{MOM}{ren_conds_4g} and $\mu = 4.3$~GeV.

 It is clear that the arguments of the $G_i$ appearing in the integrands of the SDE  
are comprised by combinations  $x_i p$ and the loop momentum $k$,
a fact that  
complicates the numerical treatment.
However, a considerable 
simplification occurs if ($\it i$) the form factor 
$G_1$ is assumed to be dominant, making negligible  
all dependence on other form-factors, in particular $G_{2,3}$, and  ($\it ii$)
the $G_1$ is assumed to display ``planar degeneracy'' at some notable level of accuracy. In such a case, one implements the substitution
\begin{align}
\overline{\fatg}_{\mu\nu\rho\sigma}^{abcd}(p_1,p_2,p_3,p_4)
    & \approx G_1(p_1,p_2,p_3,p_4)
\overline{\Gamma}_{\!0~\!\!\mu\nu\rho\sigma}^{abcd}(p_1,p_2,p_3,p_4)\,,
\label{approx}
\end{align}
and 
\begin{align}
G_1(p_1,p_2,p_3,p_4)\approx \deg(\sb^{2}) \,,
\label{planarG1}
\end{align}
where, in general,  
\begin{align}
\sb^{2}=\frac{1}{2}(p_{1}^{2}+p_{2}^{2}+p_{3}^{2}+p_{4}^{2})\,.
\label{svariable}
\end{align}

The form of 
$\deg(\bar s^2)$ will be determined by means of an iterative procedure, outlined in the next subsection.
Note that, for a given $\deg(\bar s^2)$, 
all four-gluon vertices 
in the integrands of 
\1eq{G123eqs} will be 
replaced by \2eqs{approx}{planarG1}, and evaluated at the value of  
$\bar s^2$ corresponding to each vertex. For example, the 
$\bar s^2$ assigned to the 
vertex 
$\overline{\fatg}^{admn}_{\mu\sigma\alpha\beta}(p_{1},p_{4},k,-k_{3})$ in the diagram $\left(\bar{d}_{3}\right)$ in \1eq{dis4g} is given 
by \mbox{$\sb^2= \frac{1}{2}(p_1^2+p_4^2 +k^2 +k_3^2)$}, where $k_{3}:=k+p_{1}+p_{4}$.

There are two main consequences stemming 
from the implementation of these two approximations.

First, the form factor $\T_{2}(x,p)$ vanishes identically. Indeed, we find that the diagrams  $(\bar{d}_{1,2})$ of \1eq{dis4g} cannot generate the particular color combination associated with $G_2$ [cf.~\1eq{basis_ti}], independently of the approximation used for the ghost-gluon and three-gluon vertices. On the other hand, the diagrams $(\bar{d}_{3,4})$ can contribute to $G_2$, but only if non-tree-level tensor structures of the four-gluon vertices therein are dressed. Since a nonzero $G_2$ can only arise from the effect of dressing non-tree level tensor structures, or including higher loop diagrams in the SDE, it is reasonable to expect it to be subleading in comparison to $G_{1,3}$. In particular, it is evident that, perturbatively, $G_2$ can only be nonzero at two loops or higher.

Second, the remaining equations for 
$\T_{1}(x,p)$ 
and $\T_{3}(x,p)$ {\it decouple}, 
assuming the
schematic form  
\begin{align}
   \!\!\!\! \T_{1}(x,p) &= 1 + \left[\int_{k}\!\! K_{1} + \!\!\int_{k}\!\! \deg K_{2}  + \!\!\int_{k}\!\! \degx2 K_{3}\right] -\left[ \int_k\!\!K_{1} + \!\!\int_k\!\!\deg K_{2} + \!\!\int_k\!\!\degx2 K_{3}\right]_{\substack{\!\!\!p\to \mu \\ x\to x_{0}}}\,, \label{eqG1}\\
\!\!\!\! \T_{3}(x,p) &=  \left[\int_{k}\!\! K_{4} + \!\!\int_{k}\!\! \deg K_{5}  + \!\!\int_{k}\!\! \degx2 K_{6}\right]\,,
 \label{eqG3}    
\end{align}
where the $K_i$ are the various kernels defined from the diagrams $\bar{d}_{i}^{s}(x,p)$; all functions 
other than $\deg$ have been absorbed into them. 
The term quadratic in $\deg(\bar s^2)$  
stems from graph ($d_4$); and  $x_0$ denotes 
the configuration $(0,1,1,-2)$ 
in \1eq{ren_conds_4g}.

In order to proceed with the numerical solution, the \2eqs{eqG1}{eqG3} must be passed to Euclidean space, following standard conventions (see, e.g., Eq. (5.1) of~\cite{Aguilar:2018csq}).  For the numerical treatment, the resulting integrals 
are expressed in hyper-spherical coordinates.

  Numerical integration is performed with the Gauss-Kronrod implementation of~\cite{Berntsen:1991:ADA:210232.210234}. We integrate over a grid of external squared momenta distributed logarithmically over the interval $[10^{-4},10^{4}] ~\text{GeV}^{2}$.

\subsection{Planar degeneracy as an infrared feature.}

\begin{figure}[!t]
 \includegraphics[width=0.45\linewidth]{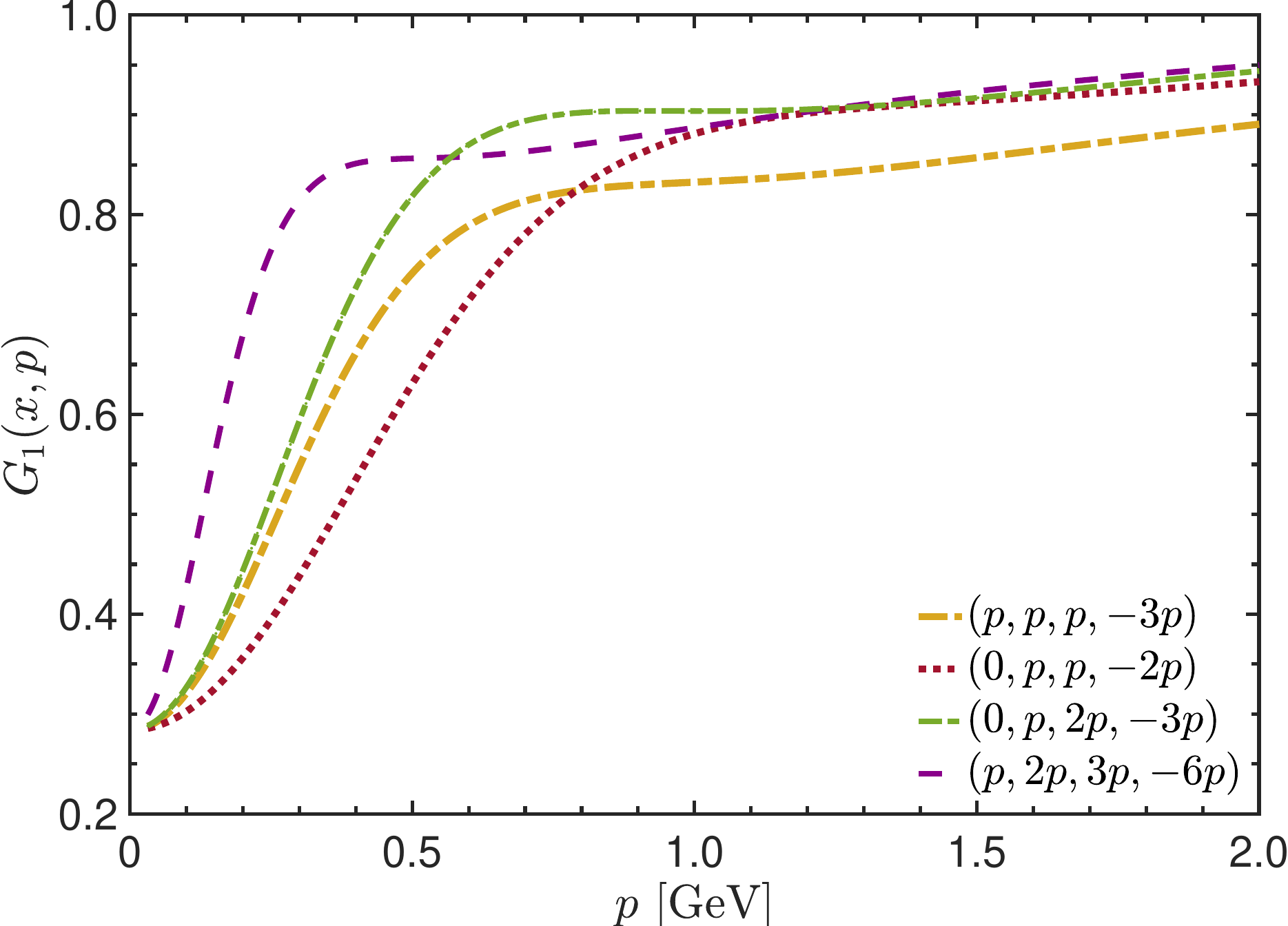}\hfil 
\includegraphics[width=0.45\linewidth]{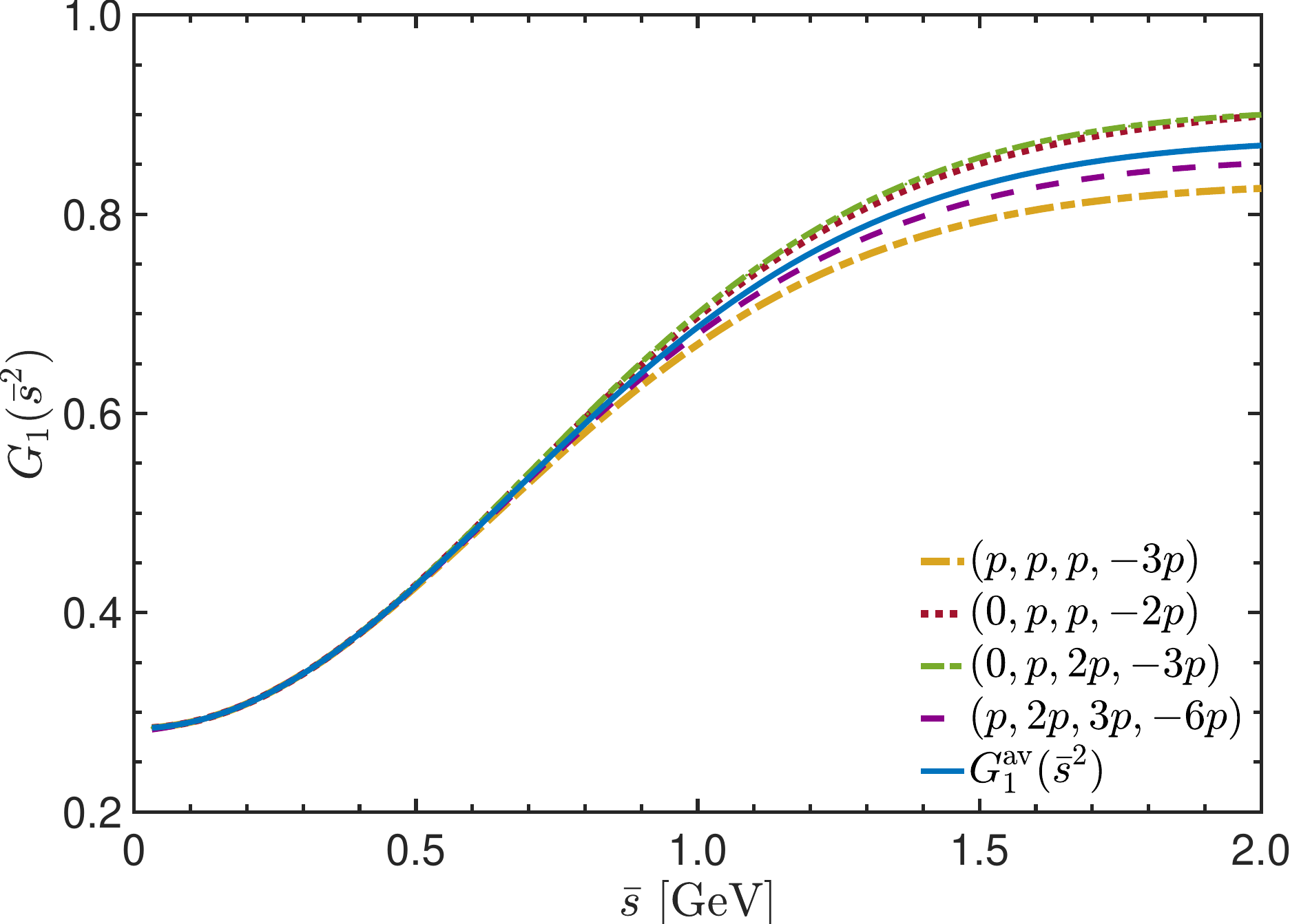}\\
\caption{Comparison of $\T_{1}(x,p)$ 
expressed as a function of the momentum $p$ (left panel) and as a function 
of the Bose symmetric variable $\bar{s}$ (right panel), for the 
 four collinear kinematic configurations, defined in \eqref{4config}.
$\T_{1}(x,p)$ was computed employing the approximation  \mbox{$\overline{\fatg} = \overline{\Gamma}_{\!0}$} on the r.h.s. of the \1eq{eqG1}. The blue curve on the right panel represents $G_1^{\rm av}(\sb^{2})$, the  average of  these four configurations in terms of  $\sb^{2}$, defined in \1eq{G1_avg_def}.}
\label{fig:Gqs2}
\end{figure}

The first step in determining 
$\deg(\sb^{2})$ is to 
set $\deg(\sb^{2}) =1$ 
everywhere on the r.h.s. of \1eq{eqG1}. 
After this substitution, 
any collinear configurations for $G_1(x,p)$ are obtained from \1eq{eqG1} by simply carrying out 
the corresponding integrations.  
Thus, we determine numerically $G_1(x,p)$  
in four collinear kinematic configurations,
whose coordinates $(x_1,x_2,x_3,x_4)$ are
given by 
\begin{align}
(1,1,1,-3) \,,\qquad 
 (0,1,1,-2)\,, \qquad  
 (0,1,2,-3) \,,\qquad 
 (1,2,3,-6)\,.
\label{4config}
\end{align}
This choice is random; there is nothing particular about these four 
cases. The results are displayed in Fig.~\ref{fig:Gqs2}: on the left panel, 
where the four configurations 
are plotted as functions of the momentum $p$, they lead to clearly distinct curves; however, when plotted in terms of the Bose symmetric variable $\bar{s}$ of \1eq{svariable},  they merge into each other in the range \mbox{$0 \leq\sb\lessapprox1$~\text{GeV}}, manifesting clearly the onset of the planar degeneracy in this region of momenta. Note that in the case of collinear momenta, \1eq{svariable} reduces to
\begin{align}
\sb^{2}=\frac{1}{2}(x_{1}^{2}+x_{2}^{2}+x_{3}^{2}+x_{4}^{2}) \, p^2 \,.
\label{scoll}
\end{align}
%

\begin{figure}[!t]
 \includegraphics[width=0.45\linewidth]{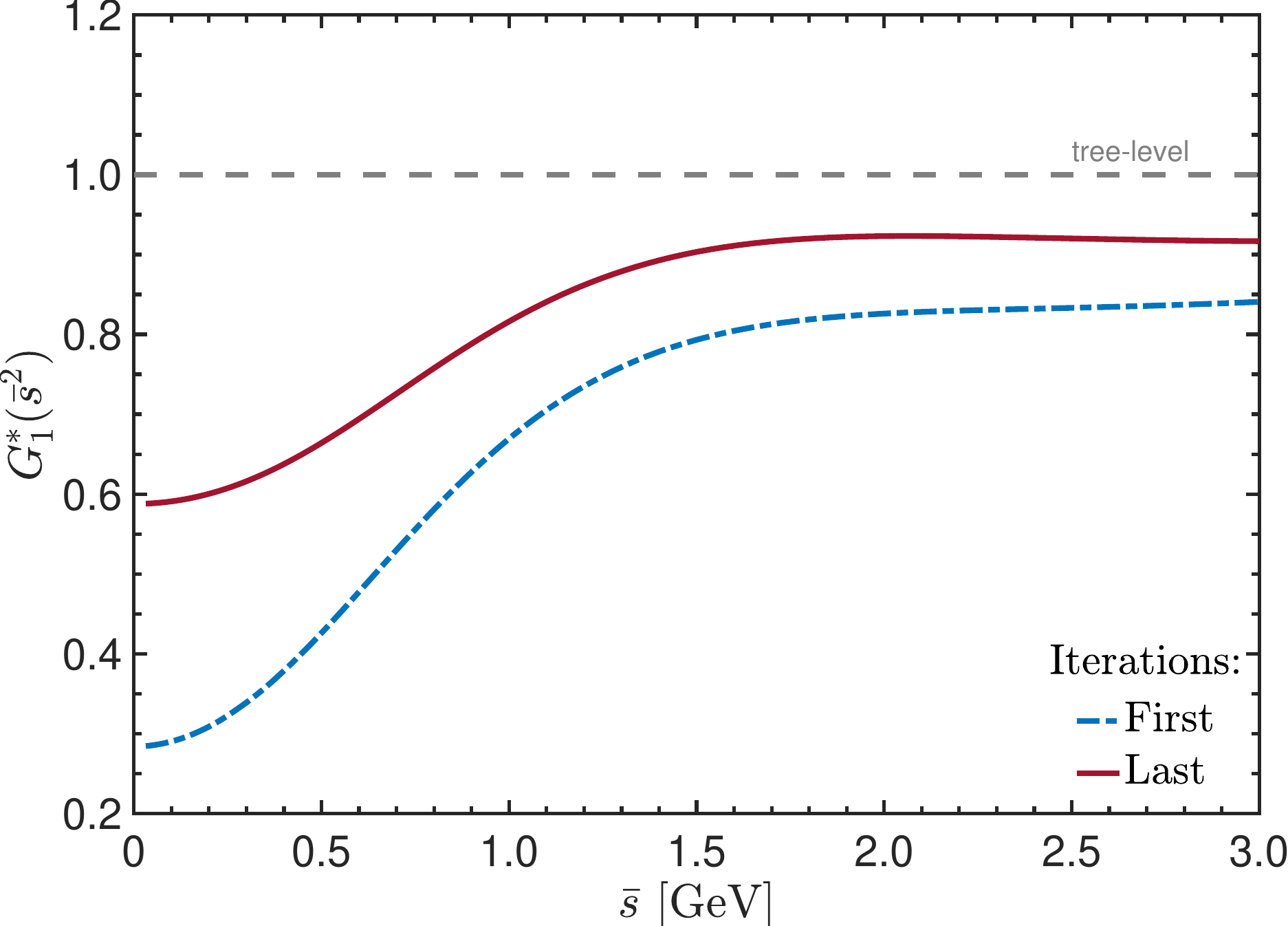}\hfil 
\includegraphics[width=0.45\linewidth]{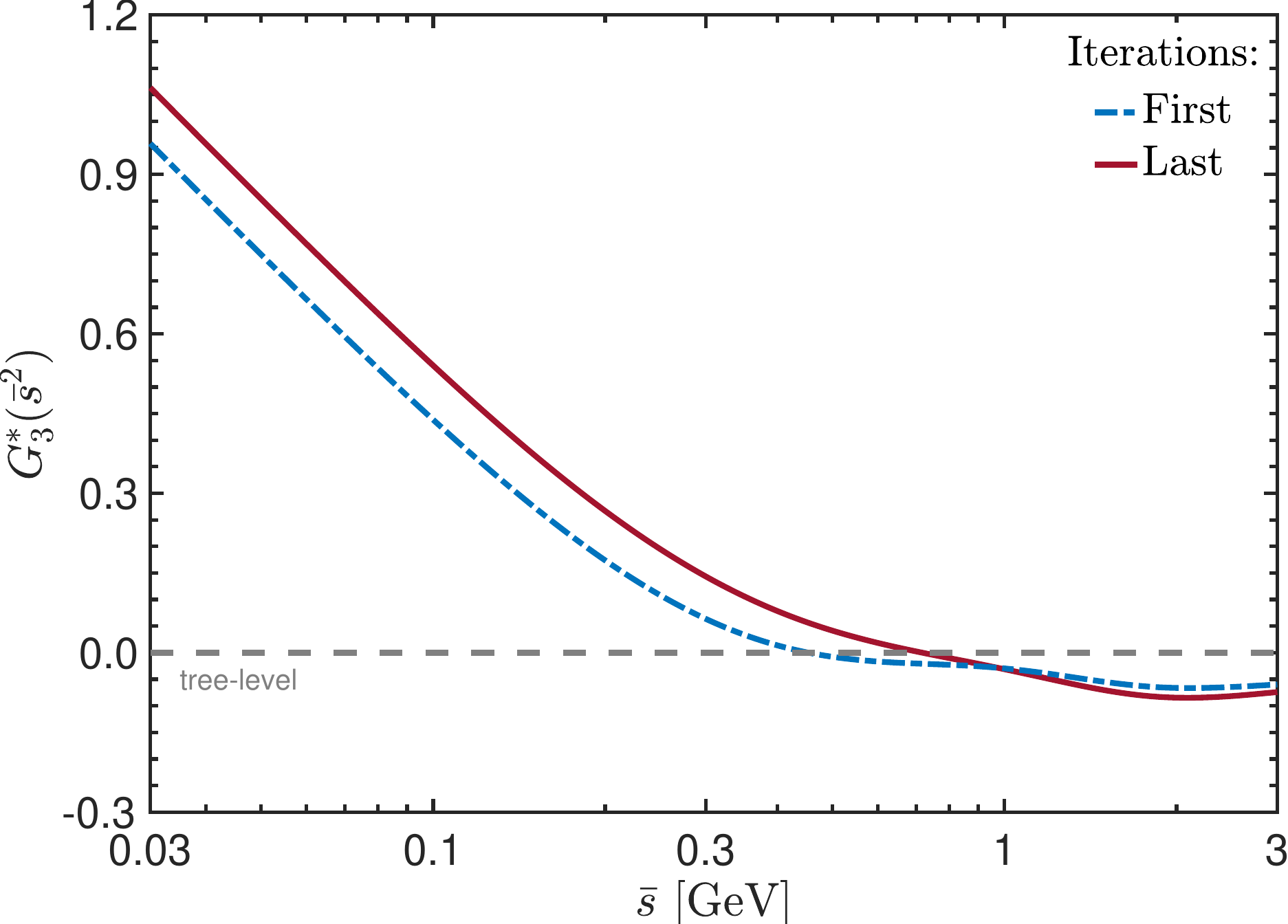}\\
\caption{The first and last iterations of $G^{*}_{1,3}(\sb^{2})$.}
\label{fig:L4g}
\end{figure}

This promising result 
motivates the iteration of the procedure, in order to obtain the 
optimal form of $\deg(\sb^{2})$. In particular, 
we determine the numerical average, 
$G_1^{\rm av}(\sb^{2})$, of the 
four configurations defined in Eq.~\eqref{4config},  \ie 
\begin{align}
    G_1^{\rm av}(\sb^{2}) = \frac{1}{4} \sum_{j=1}^{4} \T_{1}(x_{j},p)\,; \label{G1_avg_def}
\end{align}
the result of this ``averaging'' is shown 
as the blue continuous curve on the 
right panel of Fig.~\ref{fig:Gqs2}.  
Then, the second step in the iterative procedure is 
taken, by setting on the r.h.s. of \1eq{eqG1}
\begin{align}
\deg(\sb^{2}) = G_1^{\rm av}(\sb^{2}) \,;
\label{starav}
\end{align}
in this way, 
the $G_1^{\rm av}(\sb^{2})$ acts now as the new ``seed'' 
for obtaining the next generation of results for 
the same four configurations. 
These new results will then be averaged and fed back into the r.h.s. of the original set of four equations. This iterative procedure concludes when 
the relative error between two consecutive averaged solutions is smaller than $0.1\%$; typically, convergence is achieved after about 15 iterations.
In Fig.~\ref{fig:L4g}, we 
compare the $\deg(\sb^{2})$ corresponding to 
the first (blue dashed curve), and last iterations (red continuous curve).
The final result is clearly suppressed with respect to 
the initial $\deg(\sb^{2})=1$ (tree-level).

Note that this particular procedure decouples effectively  
\1eq{eqG1} from \1eq{eqG3}: the form factor $\T_3(x,p)$ may be obtained by means of a single integration, 
once the final $\deg(\sb^{2})$ has been computed from 
\1eq{eqG1} and 
substituted into 
\1eq{eqG3}. 

However, even though $G_3(x,p)$ is not subject to an iterative procedure, it is advantageous to
define a curve $G_3^{*}(\sb^{2})$, which potentially approximates the general collinear $G_3(x,p)$, and allows this form factor to be reconstructed from a single curve (see \fig{fig:SurfacesS2}). Moreover, the curve $G_3^{*}(\sb^{2})$ will serve as a reference for quantifying 
how well $G_3(x,p)$ satisfies the property of planar degeneracy. Therefore, by analogy with \2eqs{G1_avg_def}{starav}, we set 
\begin{align}
    G_3^{*}(\sb^{2}) = \frac{1}{4} \sum_{j=1}^{4} \T_{3}(x_{j},p)\,, \label{G3_star_def}
\end{align}
where the sum is over the same kinematic configurations used to define $G_1^{*}(\sb^2)$, given in Eq.~\eqref{4config}. The resulting $G_3^{*}(\sb^2)$ is shown on the right panel of \fig{fig:L4g} for the case 
$\deg(s^2) = 1$ (blue dot-dashed) and for the final $\deg(s^2)$ (red continuous).

\begin{figure}[!t]
 \includegraphics[width=0.45\linewidth]{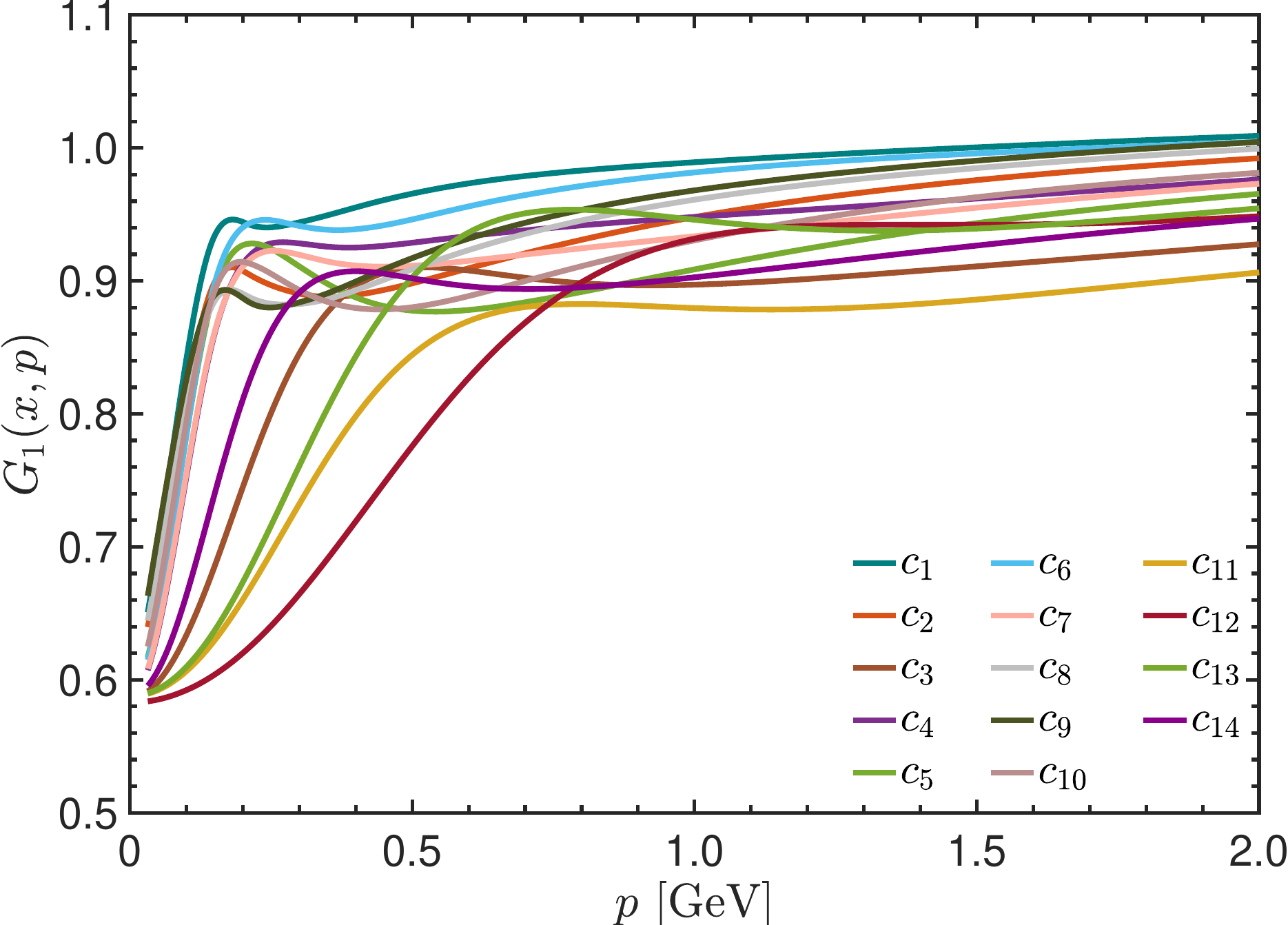}\hfil
\includegraphics[width=0.45\linewidth]{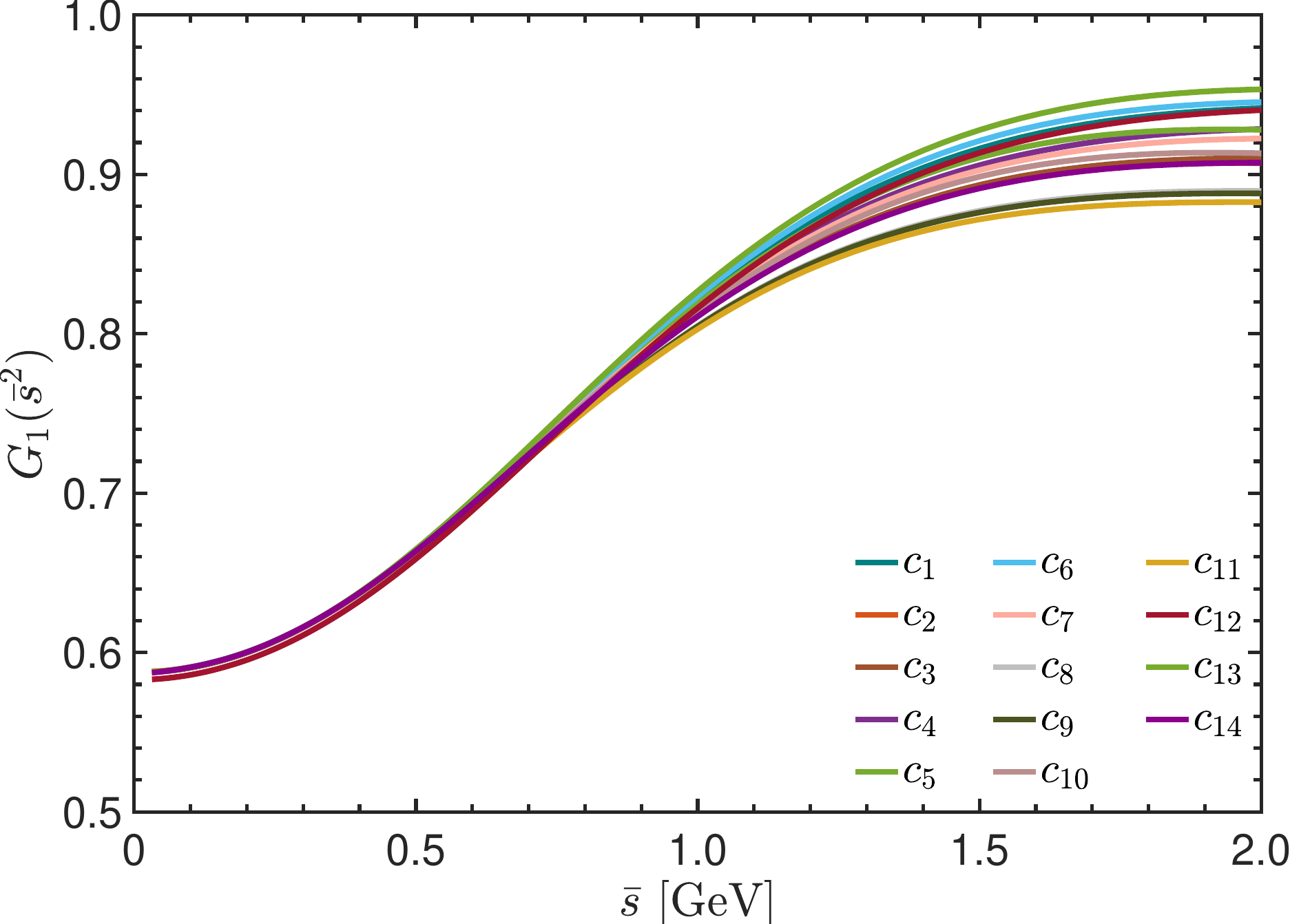}\\
\includegraphics[width=0.45\linewidth]{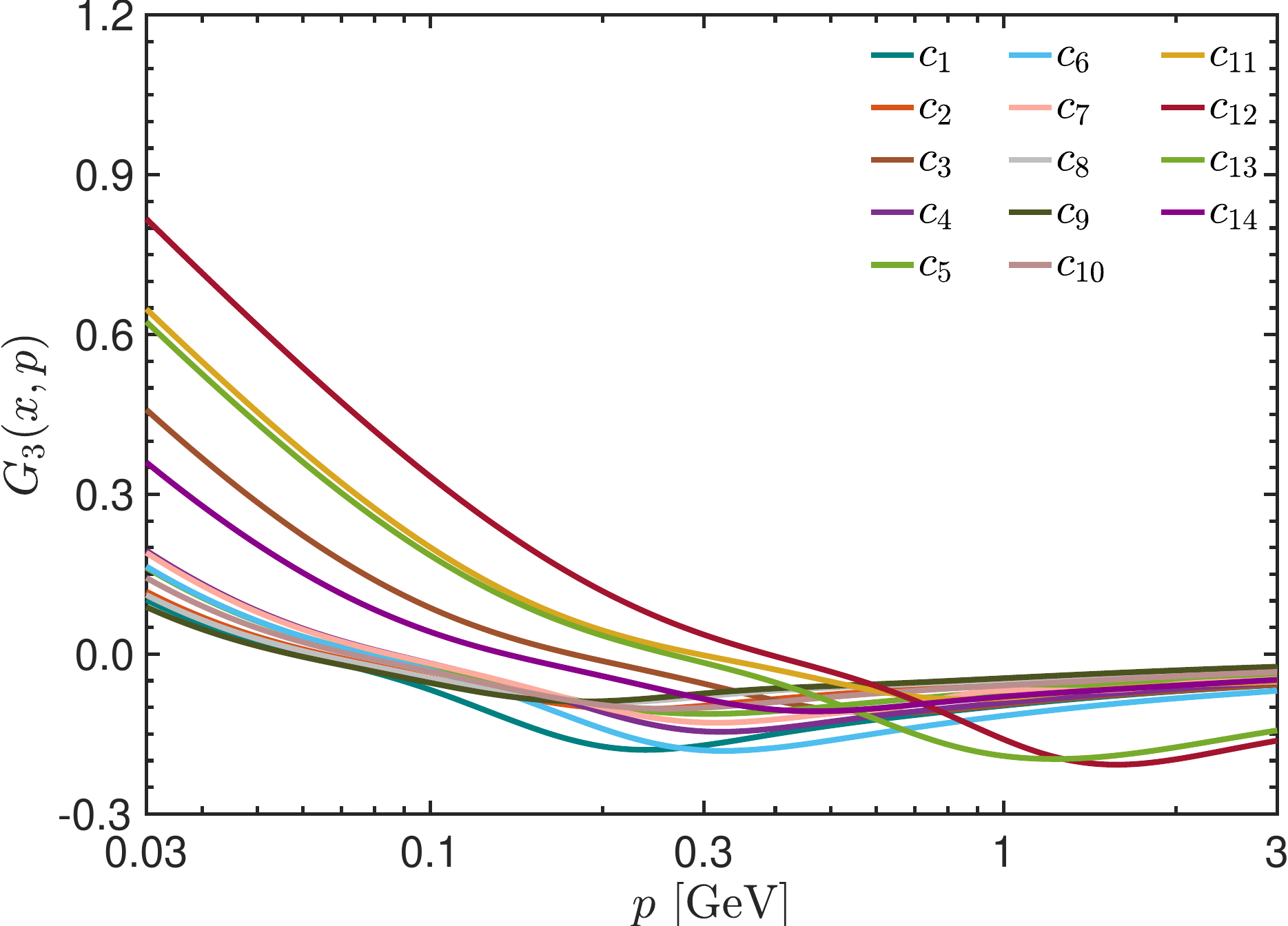}\hfil
\includegraphics[width=0.45\linewidth]{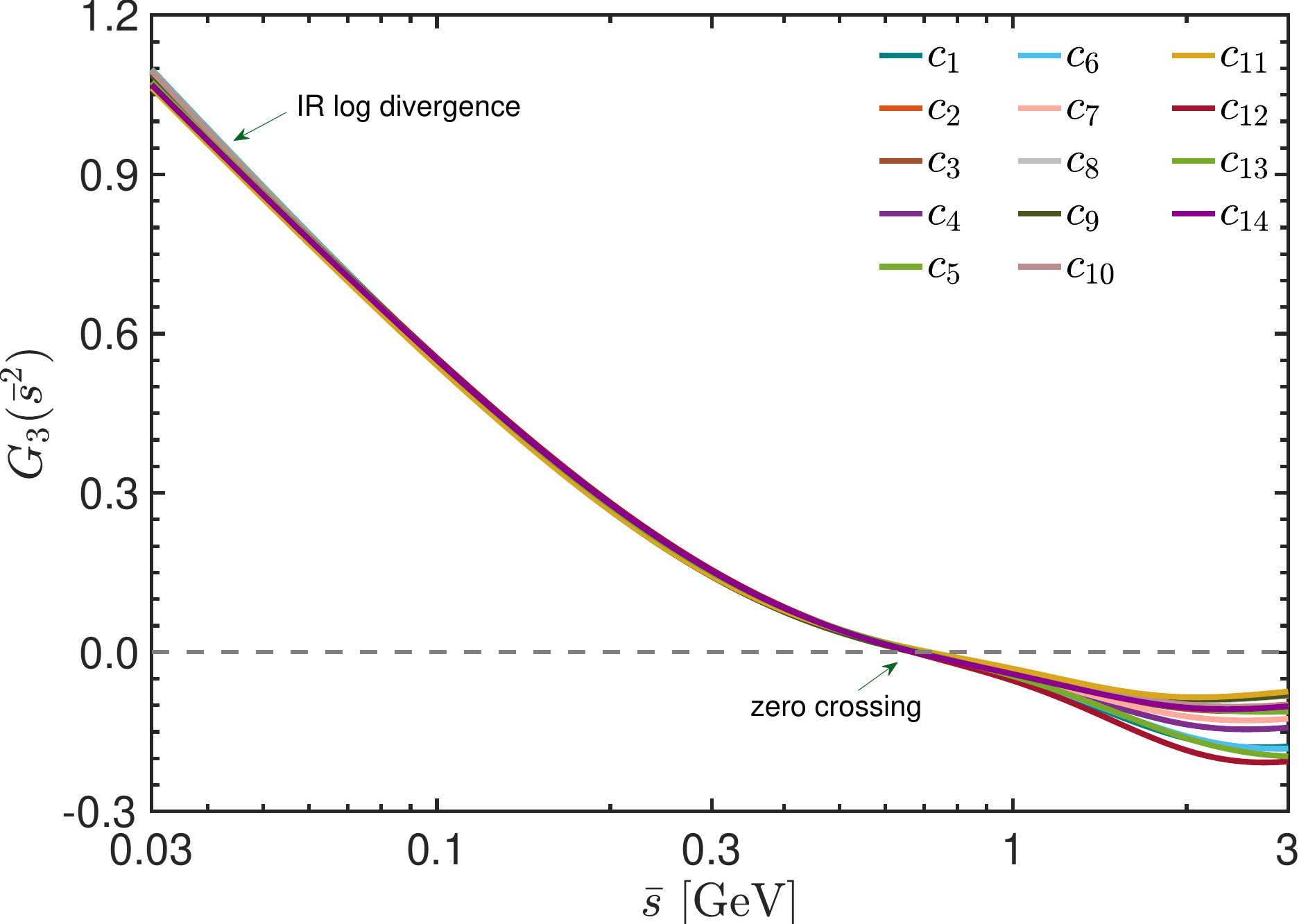}\\
\caption{ The form factors $\T_{1,3}(x,p)$ for the fourteen random collinear kinematic configurations, defined in \1eq{random_kin}, plotted as functions of momentum $p$ (left panels) and as functions of the Bose symmetric variable $\bar{s}$ (right panels). The disordered  pattern seen on the left panels is to be contrasted with the tight clustering of the curves displayed on the right ones.}
\label{fig:random_kin}
\end{figure}
As is clear from \fig{fig:L4g}, 
for $\sb \geq 0.3$~GeV, 
$\T_3^{*}(\sb^2)$ is considerably smaller than $\T_1^{*}(\sb^2)$. On the other hand, for $\sb\to 0$,  the form factor $\T_3^{*}(\sb^2)$ diverges logarithmically, whereas $\T_1^{*}(0)$ is finite. As shown in~\cite{Binosi:2014kka}, the divergence of $\T_3^{*}(\sb^2)$ results from the massless ghost loop $(d_1)$ of \fig{fig:SDE4g}, which contributes to $\T_3(x,p)$, but not to $\T_1(x,p)$. The rate of divergence can be determined by an asymptotic analysis along the lines of the one performed for the three-gluon vertex in Appendix A of~\cite{Aguilar:2023qqd}. Specifically, we obtain
\begin{align}
\lim_{p\to 0} G_{3}(x,p) = -\frac{3\alpha_{s}}{64\pi}F^{4}(0) Z_{1}^{4}\ln (p^2/\mu^2)\,,
\end{align}
where $Z_1$ is the renormalization constant of the ghost-gluon vertex. Note that $Z_1$ is finite in the Landau gauge~\cite{Taylor:1971ff,vonSmekal:1997ern,Boucaud:2011eh};
within our 
renormalization scheme, it turns out to be very close to unity, 
$Z_1 = 1.01$\,.

In order to test the reliability of the entire procedure, we compute 
$G_1$ and $G_3$ for 
fourteen  random collinear kinematics, using 
as input in \2eqs{eqG1}{eqG3} the $\deg(\sb^{2})$ obtained 
at the end of the iteration procedure (red continuous curve in \fig{fig:L4g}). 
Ten of these configurations are 
 obtained by employing a random number generator 
 for the $x_i$, while the last four correspond 
to those used earlier for the 
determination of $\deg(\sb^{2})$;  
the coordinates $(x_1,x_2,x_3,x_4)$ of all of them are listed below\footnote{Since, by Bose symmetry, $G_{1,3}(x,p)$ are invariant under a permutation of any two components $x_i$, there is no loss of generality in arranging the values of $x_1,\,x_2,\,x_3$ in ascending order; $x_4$ is fixed by momentum conservation.}:
\begin{align}
&c_1 =(0.41,5.96,6.77,-13.14)\,; \quad 
&c_6=&\,(0.23,3.78,6.17,-10.18)\,; \quad 
&c_{11} =(1,1,1,-3)\,; \nonumber  \\ 
&c_2 =(2.26,3.72,6.55,-12.53)\,; \quad 
&c_7 =&\,(1.02,3.37,5.10,-9.49)\,; \quad 
&c_{12} =(0,1,1,-2)\,; \nonumber \\
&c_3 =(0.71,1.56,2.32,-4.59)\,;  \quad
&c_8 =&\,(3.30,4.13,5.53,-12.96)\,; \quad
&c_{13}=(0,1,2,-3)\,; \nonumber \\
&c_4 =(0.74,3.96,4.63,-9.33)\,; \quad
&c_9 =&\,(4.02,4.25,6.09,-14.36)\,; \quad 
&c_{14} =(1,2,3,-6)\,. \nonumber \\
&c_5 =(1.42,2.34,6.65,-10.41)\,; \quad 
&c_{10} =&\,(2.23,2.55,6.49,-11.27)\,; \quad &
\label{random_kin}
\end{align}

In Fig.~\ref{fig:random_kin} we plot both form factors $\T_{1,3}(x,p)$, for the  configurations listed in Eq.~\eqref{random_kin}, 
as functions of the momentum $p$ (left panels) and  the Bose symmetric variable $\sb$
(right panels). The remarkable agreement between the fourteen curves when they are plotted as functions of $\sb$ is to be contrasted to the considerable disparity observed when they are plotted as functions of $p$. Thus, for $\sb < 1$~GeV, 
the approximation
\begin{align}
G_i(x,p)\approx G^{*}_i(\sb^{2}) \,, \qquad i=1,3\,,
\label{planarG3}
\end{align}
appears to be particularly robust.

To quantify the level of accuracy of  
\1eq{planarG3}
within the range $0<\sb <2$ GeV, 
we compute the percentage error defined as 
\begin{align}
    \delta_{i}(\bar{s}_j^2) = \frac{\T_{i}(\bar{s}_j^2)-G^{*}_i(\sb^{2})}{G^{*}_{i}(\bar{s}^2)}\times 100\%\,,  \quad j=1,\cdots,14\,, 
    \label{derror}
\end{align}
where $i = 1,\,3$, and $G_{i}(\bar{s}_j^2)$ 
denotes any one of 
the fourteen curves shown on the right panels of \fig{fig:random_kin}, while
$G^{*}_i(\sb^{2})$ are our reference curves,  
shown as red continuous lines in \fig{fig:L4g}.

Starting with the classical form factor, $\T_1(x,p)$, we find that for $\sb\leq 1$~GeV, the error \mbox{$\delta_{1}(\bar{s}_j^2)\leq 1.7\%$}. As can be seen already from \fig{fig:random_kin}, the relative separation between curves
of different kinematics grows as $\sb$ increases; consequently, in the range   $1\leq\sb\leq 2$~GeV,  $\delta_{1}(\bar{s}_j^2)$ increases, 
reaching the maximum value of $4.4\%$.   

For $\T_3(x,p)$, notice that, for all the kinematics shown in \fig{fig:L4g}, the curves display zero crossings at around \mbox{$\sb=0.7$~GeV}, where the relative error $\delta_3(\sb_j^2)$ becomes ill-defined. 
Away from the crossing, we find that 
$\delta_3(\sb_j^2) \leq 7.0 \%$ 
for \mbox{$\sb \leq 0.4$~GeV}, whereas for $1\leq\sb\leq 2$~GeV the $\delta_3(\sb_j^2)$ increases, reaching the 
value $\delta_3(\sb_j^2) \leq 33.7\%$;  
however, in this latter range of momenta,   
$G^{*}_{3}(\bar{s}^2)$ is itself rather small.

In addition to the 
inspection 
of discrete sets of kinematic configurations, one  may visualize the accuracy of the planar degeneracy approximation more generally through 3D plots, where the $x_i$ components vary continuously.

In \fig{fig:SurfacesS2}, we plot $G_{1,3}(x,p)$ (yellow surfaces) as functions of the momenta $x_1p$ and $x_2p$. To make this visualization possible, we reduce the number of independent variables by fixing $p_{3}=(x_1+x_2)p$. In the same figure, we map $G^{*}_{1,3}(\sb^{2})$ into the $(x_1 p$, $x_2p)$ plane, by setting $p_{3}=(x_1+x_2)p$ in \1eq{scoll}, such that 
\mbox{$\sb^{2} = 3(x_{1}p)^{2} + 3(x_{2}p)^{2} + 5 (x_{1}p)(x_{2}p)$}. With this mapping, the curves for $G_i^{*}(\sb^2)$ (red continuous in \fig{fig:L4g}) become surfaces, shown in blue in \fig{fig:SurfacesS2}. For both form factors, $G_{1,3}(x,p)$, the blue and yellow surfaces 
show a high degree of 
overlap in the infrared,  in agreement with \1eq{planarG3},  while differences become visible with increasing momenta.

\begin{figure}[t!]
 \includegraphics[scale=0.8]{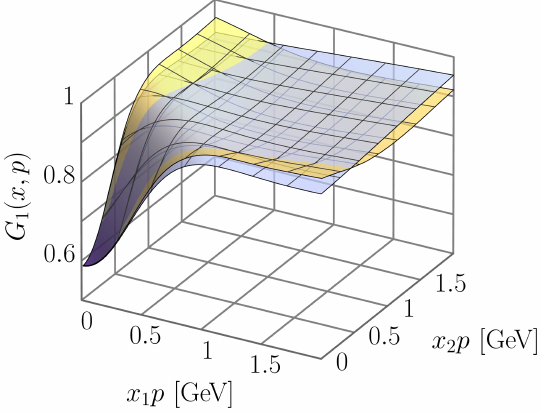}	\hfil 
 \includegraphics[scale=0.75]{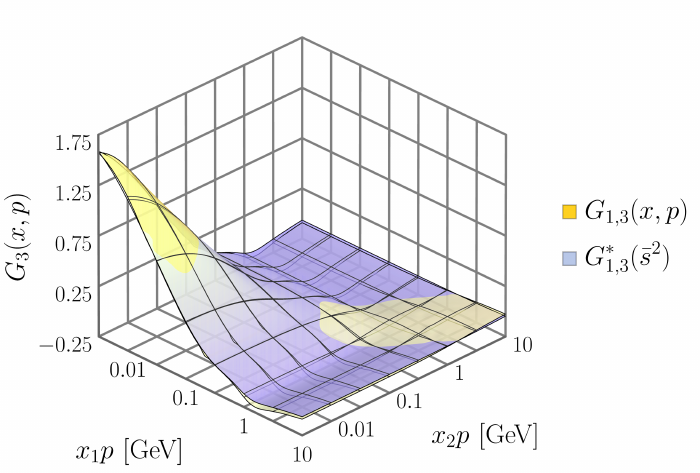}	
 \caption{Form factors $\T_{1,3}(x,p)$ (yellow) compared to the approximation $G^{*}_{1,3}(\sb^{2})$ (blue), mapped into the ($x_1p$, $x_2p$) plane using \1eq{scoll}, 
 and setting $x_{3} = x_{1} + x_{2}$. 
In both cases, the advantage of planar degeneracy becomes evident: entire surfaces are reconstructed, to a great accuracy, from the knowledge of a single curve.}
\label{fig:SurfacesS2}
\end{figure}

\begin{figure}[t!]
 \includegraphics[scale=0.77]{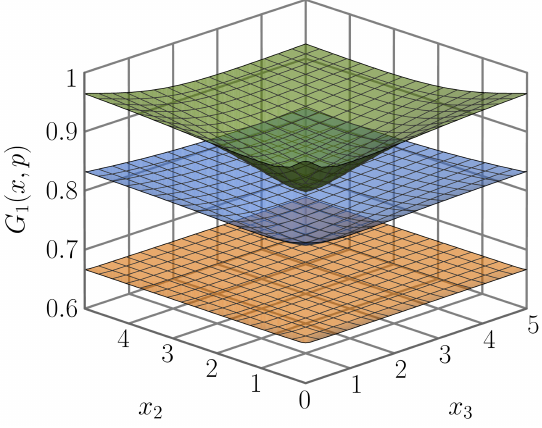} \hfil
 \includegraphics[scale=0.77]{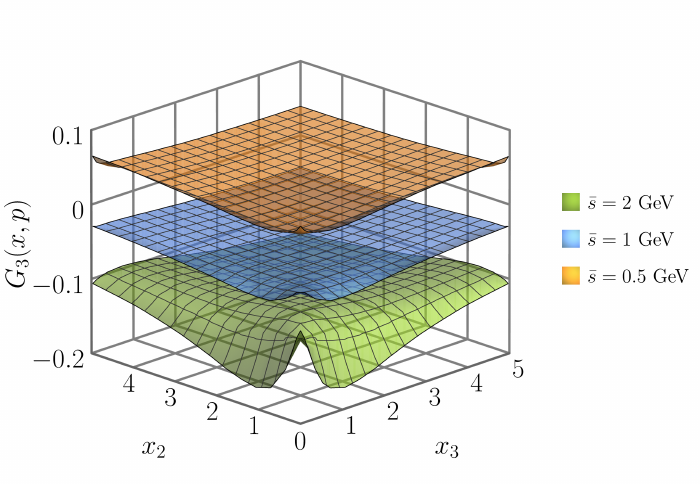}
 \caption{Form factors $\T_{1,3}(x,p)$ for \mbox{$x_1 = 1$}, arbitrary $x_2$ and $x_3$, and fixed values of \mbox{$\bar{s} = 0.5$~GeV} (brown), \mbox{$\bar{s} = 1$~GeV} (blue), and \mbox{$\bar{s} = 2$~GeV} (green). For \mbox{$\bar{s} = 1$~GeV} and \mbox{$\bar{s} = 2$~GeV}, both $\T_{1,3}(x,p)$ are practically constant, \ie independent of $x_i$.}
\label{fig:lasagna}
\end{figure}

Let us finally point out that, if planar degeneracy were an exact property of the $\T_i(x,p)$, there should be no dependence on $x$ for  fixed $\sb$; 
instead, for an approximate planar degeneracy, the $\T_i(x,p)$ should show a mild dependence on $x$. The residual dependences of the $\T_i(x,p)$ on $x$ can be visualized by plotting surfaces corresponding to fixed values of $\sb$; 
their deviation from the absolute flatness 
will be then a measure of the 
exactness of the planar degeneracy.

To this end, first note that for a given vector $x$, the value of $p$ that gives rise to a given value of $\sb$ can be uniquely determined by inverting \1eq{scoll}. Specifically,
\be 
p = \sb\left(x_1^2 + x_2^2 + x_3^2 + x_1 x_2 + x_2 x_3 + x_3 x_1\right)^{-1/2} \,, \label{p_from_s}
\ee
where $x_4$ has been eliminated 
by using momentum conservation. 
Hence, we can determine how $G_1(x,p)$ varies as a function of $x_1$, $x_2$ and $x_3$, for fixed $\sb$: 
from an available array of 
values for $G_1(x,p)$, we pick the
one value that corresponds to the unique $p$ obtained from
\1eq{p_from_s}, once an $\sb$ and a set of 
$(x_1,x_2,x_3)$ have been substituted in it.

In \fig{fig:lasagna} 
we first set $x_1=1$, and then plot $\T_{1,3}(x,p)$ for fixed values of $\sb$ and general values of $x_2$ and $x_3$.  
Specifically, we show surfaces corresponding to \mbox{$\sb = 0.5$~GeV} (brown), \mbox{$\sb = 1$~GeV} (blue), and \mbox{$\sb = 2$~GeV} (green). In both panels, it is clear that for \mbox{$\bar{s} = 0.5$~GeV} and  \mbox{$\bar{s} = 1$~GeV}, the $\T_{1,3}(x,p)$ are
almost perfectly constant, \ie nearly independent of $x_2$ and $x_3$. Hence, we confirm once more the validity of the planar degeneracy
property expressed by~\2eqs{planarG1}{planarG3} in the infrared.

For $\sb = 2$~GeV, the surfaces of $\T_{1,3}(x,p)$ in \fig{fig:lasagna} begin to deviate
from perfect flatness, in line with our previous observations that planar degeneracy breaks down with increasing $\sb$. However, even in this case, the surfaces of $\T_{1,3}(x,p)$ are rather flat for most values of $x_2$ and $x_3$. The only exceptions occur near the borders of the plot, 
when at least one of the $x_i$ is small, with particularly stronger dependences on $x$ when both $x_2,\,x_3 \approx 0$.

\section{Comparison with previous works: Impact of the three-gluon vertex}
\label{sec:comparison}

In this section, we compare our results with previous continuum studies of the four-gluon vertex. We recall that in the present work we employed a 4PI-based truncation of the SDE, given by \1eq{fig:SDE4g}, where all vertices in the one-loop diagrams appear dressed. On the other hand, in~\cite{Binosi:2014kka} all vertices on the r.h.s. of the SDE are set to tree-level, whereas in~\cite{Cyrol:2014kca,Huber:2018ned} one vertex per diagram appears at tree-level. As such, the comparison of these results to ours allows us to assess the effect that the dressing
of the vertices, and especially of the three-gluon vertex, has  
on the four-gluon SDE.

For this task, we will focus on the configuration $(p,p,p,-3p)$, corresponding to $c_{11}$ of \fig{fig:random_kin}, which has also been studied in~\cite{Binosi:2014kka,Cyrol:2014kca,Huber:2018ned}; for simplicity, we limit our discussion to the classical form factor
$G_1$.

In this kinematic limit, all tensors $t_{i,\mu\nu\rho\sigma}^{abcd} (x,p)$ for $i \geq 4$  vanish\footnote{This is easily established with the formalism of Appendix~\ref{basis_col}, by noting that $\overline{\mathcal{D}} = 0$ and $\overline{\mathcal{T}}\star\overline{\mathcal{T}} = 0$ in this configuration, and that only $t_{1,2,3}$ in \1eq{tibasis} are independent of both $\overline{\mathcal{D}}$ and  $\overline{\mathcal{T}}\star\overline{\mathcal{T}}$.}. Then, \1eq{basis} reduces to
\be 
\overline{\fatg}^{abcd}_{\mu\nu\rho\sigma}(x,p) = \sum_{i = 1}^{3} \T_i (x,p)  \, t_{i,\mu\nu\rho\sigma}^{abcd}(x,p) \,, 
 \qquad x = (1,1,1,-3) \,.
\ee
Then, it is straightforward to show that the form factor denoted by $V_{\Gamma^{(0)}}(p^2)$ in Eq.~(4.17) of~\cite{Binosi:2014kka}, and shown in Fig.~5 of that reference, corresponds precisely to $G_1(p,p,p,-3p)$.

Similarly, the projection 
\be
D^{\srm{4g}}(p_1,p_2,p_3,p_3) := \frac{\gbar^{0}(p_1,p_2,p_3,p_4)\odot \overline{\fatg}(p_1,p_2,p_3,p_4)}{\gbar^{0}(p_1,p_2,p_3,p_4)\odot\gbar^{0}(p_1,p_2,p_3,p_4)}\,,
\ee
computed in~\cite{Cyrol:2014kca}, reduces to $D^{\srm{4g}}(p,p,p,-3p) = G_1(p,p,p,-3p) + (2/3) G_2(p,p,p,-3p)$. Since under our approximations $G_2 = 0$, we can set $D^{\srm{4g}}(p,p,p,-3p) = G_1(p,p,p,-3p)$, and compare our result to that shown on the upper left panel of Fig.~8 of~\cite{Cyrol:2014kca}.

Note that~\cite{Binosi:2014kka,Cyrol:2014kca} employed renormalization schemes different to the one used in the present work, defined by the prescription in \2eqs{MOM}{ren_conds_4g}. Hence, to perform a meaningful comparison, we rescale all results to be equal to $1$ at $\mu = 4.3$~GeV, \ie
\begin{align} 
G_1(p,p,p,-3p)\to&\,\, \frac{G_1(p,p,p,-3p)}{G_1(\mu,\mu,\mu,-3\mu)} \,.
\end{align}
%

\begin{figure}[t!]
\includegraphics[width=0.6\linewidth]
{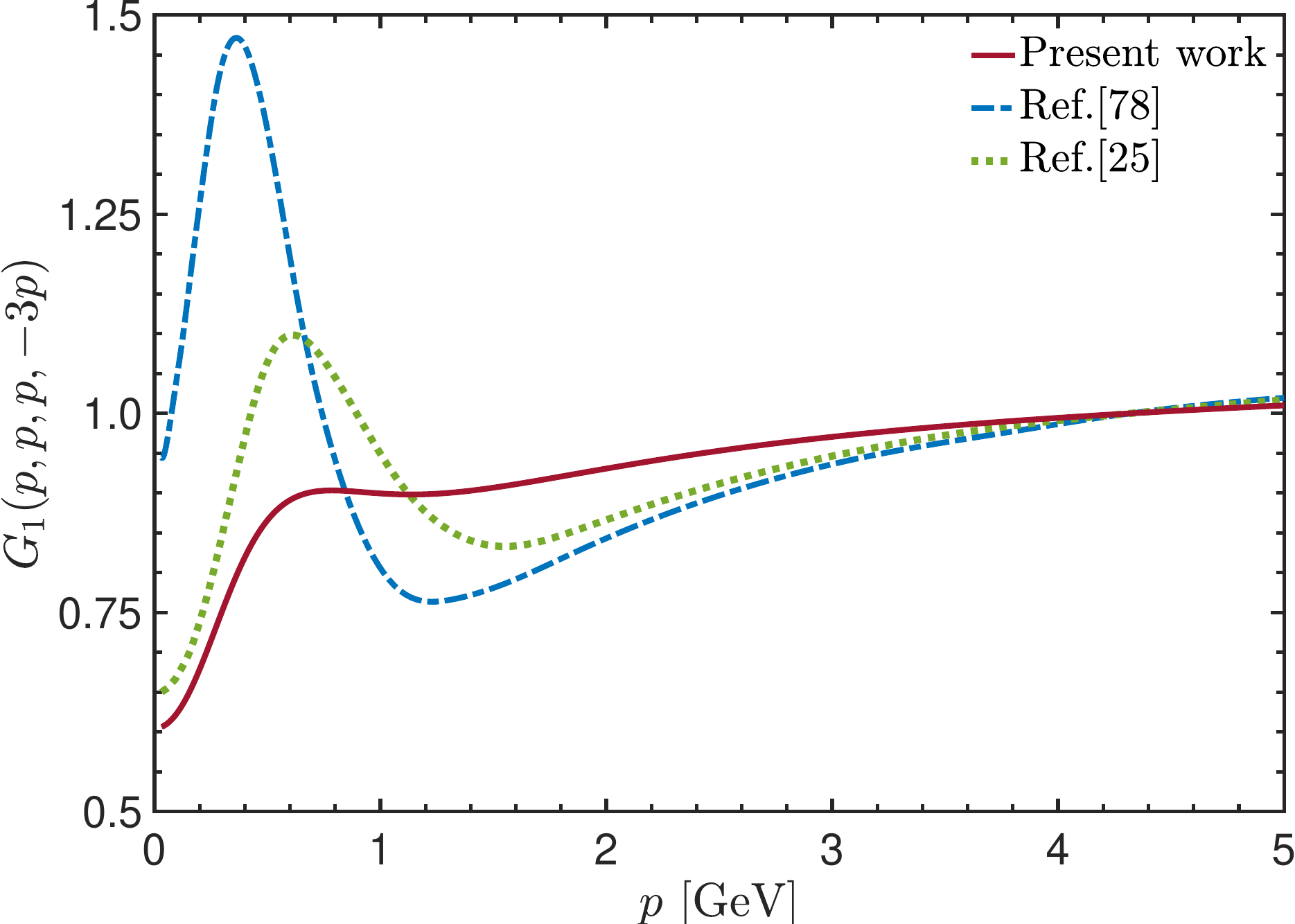} 
 \caption{Comparison of our results (red continuous line) for the configuration $c_{11}$ [\1eq{random_kin}]  with 
previous results, obtained in~\cite{Binosi:2014kka} (blue dashed), and~\cite{Cyrol:2014kca} (green dotted).}
\label{fig:previous_works}
\end{figure}

In \fig{fig:previous_works} we show our result for $G_1(p,p,p,-3p)$ as a continuous red line, while the result of~\cite{Binosi:2014kka} is shown in blue dot-dashed, and that of~\cite{Cyrol:2014kca} corresponds to the green dotted curve. The most prominent difference is seen for $p<1.5$~GeV. In this region, the results of~\cite{Binosi:2014kka,Cyrol:2014kca} display a large peak, which is nearly absent in our $G_1(p,p,p,-3p)$. Indeed, in this range of momenta, our result has a local maximum of $0.9$ at $p = 0.8$~GeV. In contrast, the result of~\cite{Binosi:2014kka} is strongly enhanced in the infrared, reaching a maximum of $1.5$ at $0.4$~GeV, while~\cite{Cyrol:2014kca} obtains a smaller maximum of~$1.1$ at $0.6$~GeV, and becomes very similar to ours for $p\lessapprox 0.4$~GeV.

To understand the origin of these differences in the  infrared behavior, we have computed the individual contributions of each diagram of \fig{fig:SDE4g}, and compared to the Fig.~5 of~\cite{Binosi:2014kka} and Fig.~8 of~\cite{Cyrol:2014kca}. We find that, 
in the infrared, 
the diagrams $(d_2)$ and $(d_3)$ contribute positively to $G_1(p,p,p,-3p)$, in agreement with~\cite{Binosi:2014kka,Cyrol:2014kca}, but vary considerably in size within the different truncations.

Then, we note that both of these diagrams contain three-gluon vertices. As has been firmly established by lattice~\cite{Cucchieri:2008qm,Athenodorou:2016oyh,Duarte:2016ieu,Boucaud:2017obn,Sternbeck:2017ntv,Aguilar:2019uob,Aguilar:2021lke} and continuum studies~\cite{Aguilar:2013vaa,Huber:2012kd,Blum:2014gna,Eichmann:2014xya,Williams:2015cvx,Ferreira:2023fva}, the three-gluon vertex is strongly suppressed in the infrared in comparison to its tree-level value. Hence, in our treatment, the diagrams $(d_2)$ and $(d_3)$ yield minimal contributions, due to the suppressing effect of dressing the three-gluon form factor $\Ls(s^2)$ [cf.~\1eq{compact}]. Consequently, the shape of our $G_1$ is dominated by the tree-level and the negative $(d_4)$ diagram.

In contrast, in~\cite{Binosi:2014kka}, where all vertices on the r.h.s. of the SDE are at tree-level, $(d_2)$ and $(d_3)$ give large contributions, giving rise to the peak seen in \fig{fig:previous_works}. Finally, the truncation of \cite{Cyrol:2014kca} yields a result between ours and that of~\cite{Binosi:2014kka}, since there is only one tree-level vertex per diagram. 

From these observations we conclude that the approximation employed for the three-gluon vertex has a profound impact on SDE analyses of the four-gluon vertex.

Finally, we point out that the four-gluon vertex has also been studied in the context of scaling solutions in~\cite{Kellermann:2008iw,Cyrol:2014kca,Huber:2018ned}. In this case, the classical form factor displays infrared divergences that are absent in our framework.

\section{Ultraviolet Finiteness of the form factors \texorpdfstring{$G_1$}{G1} and \texorpdfstring{$G_3$}{G3}}
\label{sec:Gi_finite}

Even though 
the numerical evaluation of the 
SDEs presented in the previous section gave rise 
to results 
that are free of ultraviolet 
divergences, it would be clearly 
advantageous to devise an analytic
demonstration of this fact.
In this section, we 
show that when the SDE undergoes the 
renormalization procedure outlined in Sec.~\ref{sec:sde}, 
and under the assumptions  discussed in Sec.~\ref{sec:numA}, 
one obtains ultraviolet finite 
form factors $G_1$ and $G_3$. In particular, 
$G_1$ becomes finite after a single 
subtraction, while $G_3$ turns out 
to be 
automatically free of ultraviolet 
divergences, 
exactly as expected.

We begin by writing the 
SDEs in \1eq{G123eqs} in   
Euclidean space using hyperspherical 
coordinates; then, integrating  
over the angle between $k$ and $p$, 
and setting $y := k^{2}$,  we arrive at 
\begin{align}
    G_i(x,p) =
    \int_{0}^{\LambdaUV^2}\!\!\!\!\!\!\!
    dy \,f_i(y,p^2,x)\,, \qquad i=1,3 \,,
\label{Gint}
\end{align}
where $\LambdaUV^2$ is an ultraviolet cutoff,
and $f_i(y,p^2,x)$ denotes the sum of the integrands 
of the projected diagrams. The closed expressions for the $f_i(x,p)$ are rather complicated;
however, for the purposes of renormalization,
only their behavior for asymptotic $y$ is 
relevant.

At this point we introduce an arbitrary 
mass scale $M$, with $p^2, \mu^2 \ll M^2$, 
and split the integrals in 
\1eq{Gint} into a finite 
and a potentially divergent part, \ie 
\begin{align}
    \!\!G_{i}(x,p) =
    \int_{0}^{M^2}\!\!\!\!\!\!\!\! dy \,f_i(y,p^{2},x) 
    \,+\, \underbrace{\int_{M^2}^{\LambdaUV^2}\!\!\!\!\!\!\!
    dy \,f_i(y,p^{2},x)}_{\rm potentially \,\, divergent} \,. 
\label{divergent}
\end{align}
The type of divergence contained 
in the second integral may be 
established by performing a Taylor expansion of $f_i(y,p^2,x)$ around \mbox{$p^2/M^2=0$},  
\begin{align}
    f_i(y, p^{2}, x) &= 
    f_i(y) + 
    p^{2}\left[\frac{\partial f_i(y, p^{2}, x)} {\partial p^{2}}\right]_{\!p^2= 0} + \ldots\,,
\label{derivative}
\end{align}
where the ellipsis denotes higher derivatives.
Note that, upon setting $p^2=0$, 
all reference to the 
parameter $x :=(x_1,x_2,x_3,x_4)$ 
drops out from $f_i(y)$.

Evidently, for the renormalization to go through, the only potential divergence must 
be isolated in the term $f_i(y)$, whereas 
the terms with the derivatives 
must yield finite integrals when substituted into \1eq{divergent}.

To check if this indeed so, 
let us first focus on $f_3(y)$; 
a detailed calculation reveals that 
it consists of four contributions,
$I_{i}(y)$, one  
from each diagram $(\bar{d}_{i}^{s})$ 
$[i=1,2,3,4]$, namely 
\begin{align}
\label{UVints}
    f_{3}(y) &=  
     \frac{3\alpha_{s}}{64\pi} \left[I_{1}(y) -24 \, I_{2}(y)
    + 40 \, I_{3}(y) -17 \, I_{4}(y) \right]\,,
\end{align}
where
\bea
\label{G3divs}
    &I_1(y) = y^{-1} B_{1}^{4}(y,y,\pi) F^{4}(y)\,;  \qquad\quad 
    &I_2(y) =  y^{3}\Delta^{4}(y)\Ls^{4}(y)\,;\nonumber\\
   &I_3(y) =y^{2}\Delta^{3}(y)\Ls^{2}(y) \deg(y)\,;  \qquad \quad
    &I_4(y) = y\Delta^{2}(y) \degx2(y)\,.
\eea

To further evaluate the $I_i(y)$, 
we assume that, in the limit $y\to \infty$, the functions comprising them 
reproduce their one-loop resummed forms~\cite{vonSmekal:1997ern,Fischer:2002eq,Pennington:2011xs,Huber:2018ned} \ie  
\begin{align}
    \label{AnDim}
    F(y)&\sim \Luv^{-9/44}(y)\,, &&
    \Delta(y)\sim y^{-1}\Luv^{-13/22}(y)\,,&&
    B_{1}(y,y,\pi)\sim 1\,, \nonumber\\
    \Ls(y)&\sim \Luv^{17/44}(y)\,, &&
    \deg(y)\sim \Luv^{2/11}(y)\,, && \Luv(y) := c\,\ln(y/\LambdaMOM^{2}) \,,
\end{align}
where $c = 1/\ln(\mu^2/\LambdaMOM^{2})$, and $\LambdaMOM$ is the QCD mass-scale in the MOM scheme. 
Note that this assumption 
is built into the 
fits employed for all quantities 
that serve as inputs for the SDE: 
indeed, the expressions for $F(y)$, $\Delta(y)$, and $\Ls(y)$
given respectively in Eqs.~(C6), (C11) and~(C12) of~\cite{Aguilar:2021uwa} are designed to 
capture the correct one-loop resummed anomalous dimension, while the numerical results in Fig.~13 of \cite{Ferreira:2023fva} satisfy $B_1(k,-k,0)\to 1$ at large $k$.

The anomalous dimension of $G_1$ can be obtained by standard renormalization group analysis~\cite{Altarelli:1981ax} from the one-loop form of the renormalization constant $Z_4$, given in \1eq{Z4_4g}. Specifically, the exponent $d_{\srm{4g}}$ of the logarithm in $G_1$ is given by $d_{\srm{4g}} = \gamma_{\srm{4g}}^{(0)}/\beta_0$, where \mbox{$\beta_0 = 11/(4\pi)$} and $\gamma_{\srm{4g}}^{(0)}$ are the lowest order coefficients of the beta function, $\beta(\alpha_s)$, and the four-gluon anomalous dimension, $\gamma_{\srm{4g}}(\alpha_s)$, respectively, \ie
\be 
\beta(\alpha) = \frac{\partial \alpha}{\partial \ln\mu^2} = - \beta_0 \alpha_s + \ldots \,, \qquad  \gamma_{\srm{4g}}(\alpha_s) = \frac{1}{Z_4}\frac{\partial Z_4 }{\partial \ln \mu^2} = - \gamma_{\srm{4g}}^{(0)} \alpha_s + \ldots \,,
\ee
with ellipses denoting higher powers of $\alpha_s$. Then, \1eq{Z4_4g} entails $\gamma_{\srm{4g}}^{(0)} = 1/(2\pi)$ and, hence \mbox{$d_{\srm{4g}} = 2/11$}.

Using \1eq{AnDim} in \1eq{G3divs}, 
we find that all $I_i(y)$ are equal, 
namely 
\be
I_i(y) = \frac{1}{c^{9/11} y \ln^{9/11} (y/\LambdaMOM^{2})}
:= I(y) \,, \qquad \forall i \,.
\label{Alli}
\ee
Note that the individual contributions 
from each diagram diverge, since 
the integral
\begin{align}
{\cal I} := \int_{\!M^{2}}^{\LambdaUV^{2}} \!\!dy \,I(y) =
    \frac{11}{2}c^{-9/11}\left[ \ln^{2/11} (\LambdaUV^2/\LambdaMOM^{2}) -\ln^{2/11} (M^2/\LambdaMOM^{2})\right]\,,
\end{align}
diverges as $\LambdaUV^2 \to\infty$.
However, the substitution of \1eq{Alli} into \1eq{UVints} reveals that 
\begin{align}
\label{f3_finite}
    f_{3}(y) &=  
     \frac{3\alpha_{s}}{64\pi} \left[1-24
    + 40 -17 \right] I(y) = 0 \,,
\end{align}
namely that the full asymptotic integrand 
$f_3(y)$ vanishes before carrying out any integration.

Turning to the form factor $G_1$, note that 
only $(\bar{d}_{3}^{s})$ and 
$(\bar{d}_{4}^{s})$ contribute to it, yielding 
\be
\label{G1UVints}
    f_{1} (y) =  
     \frac{\alpha_{s}}{2\pi} \left[I_3(y) - 2 I_4(y)\right]
 =  - \frac{\alpha_{s}}{2\pi} I(y) \,,    
\ee
which upon integration yields 
the divergent contribution of $G_1$, 
given by the constant
\be
G_1^{\rm div} = -\frac{\alpha_{s}}{2\pi} \, {\cal I}
\,.
\label{G1div}
\ee
Clearly, the subtraction prescribed by the 
first relation in \1eq{G123eqs} will cancel 
the divergence in \1eq{G1div}, giving rise to 
a finite $G_1$. Specifically,
the exact same steps leading from 
\1eq{Gint} to \1eq{G1div} may be repeated 
for $f_1(y,\mu^2, x_0)$, expanding around 
$\mu^2/M^2=0$; $x_0$ denotes 
the reference configuration $(0,1,1,-2)$ 
in \1eq{ren_conds_4g}.                      

\begin{figure}[t!]
\includegraphics[width=0.6\linewidth]{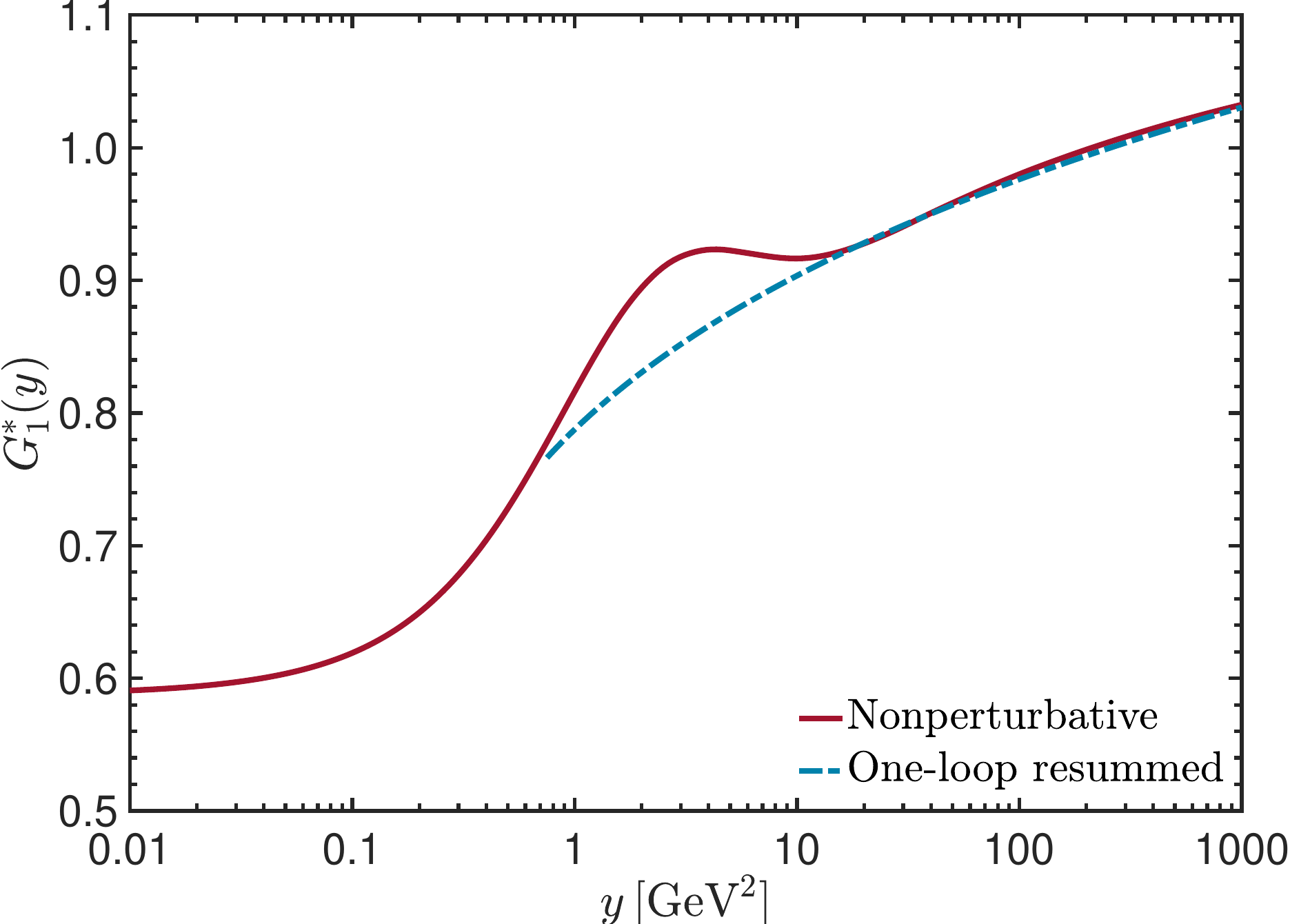}
\caption{Comparison of the nonperturbative solution for 
$\deg(y)$ with its one-loop resummed behavior given by
\1eq{AnDim}.}
\label{fig:G1pert}
\end{figure}

We end this section by confirming that 
the higher derivative terms in \1eq{derivative} are 
ultraviolet finite, and do not interfere with 
renormalization. In particular, with the aid 
of \1eq{AnDim}, we obtain 
\begin{align}
\left[\frac{\partial f_i(y, p^{2}, x_{i})} {\partial p^{2}}\right]_{\!p^2= 0} = 
   \frac{ a_{i} + b_{i}\log(y/\LambdaMOM^{2}) + c_{i}\log^{2}(y/\LambdaMOM^{2})}{y^2 \log^{31/11}(y/\LambdaMOM^{2})} + \ldots\,, \label{subder}
\end{align}
for constants $a_{i}$, $b_{i}$ and $c_{i}$ that depend 
in general on $x$, and  
the ellipsis denotes terms that are subleading for large $y$. Integrating the r.h.s. of 
\1eq{subder} over $y$ yields a convergent result; 
keeping higher order derivatives 
in the expansion of \1eq{derivative}
leads to even faster converging integrals.

It is important to emphasize that, 
for asymptotic values of the momentum, our numerical 
solutions for $\deg$ are completely compatible 
with the anomalous dimension of 
$2/11$, quoted in \1eq{AnDim}; 
this is shown in Fig~\ref{fig:G1pert}.

\section{Four-gluon effective charge}
\label{sec:charge}

Typically, the quantitative 
description of the strength of a 
given interaction proceeds through 
the construction of the corresponding 
\emph{renormalization group invariant} (RGI) effective charge~\cite{Binosi:2002vk,
Alkofer:2004it,Kellermann:2008iw,Aguilar:2009nf,Aguilar:2010gm,Boucaud:2011ug,Deur:2016tte,Binosi:2016nme,Cui:2019dwv,Deur:2023dzc}. 
In the case of the four-gluon 
interaction, the 
appropriate RGI combination
is obtained with the aid of 
the relation 
\mbox{$Z_g^{-1} = Z_4^{-1/2} Z_A$}
[see \1eq{eq:sti_renorm}]. Specifically, 
introducing the gluon dressing 
function ${\cal Z}(p^2) := p^2\Delta(p^2)$,
it is straightforward to establish that 
the combination 
\be
\alpha_{\srm{4g}}(x,p) =\, \alpha_{s}
\T_1 (x,p) {\cal Z}^2(p^2)
= \alpha^{\s R}_{s}(\mu^2) 
\T_1^{\s R} (x,p, \mu) {\cal Z}^2_{\s R}(p^2,\mu^2) \,,
\label{4geff}
\ee 
retains the same form before and after renormalization,
and thus, does not depend on the renormalization point 
$\mu$, event though the individual components are 
$\mu$-dependent. Then, the quantity 
$\alpha_{\srm{4g}}(x,p)$ will be constructed using inputs renormalized at a given $\mu$, which we choose 
$\mu = 4.3$ GeV;  evidently, any other choice of $\mu$ 
would be equally good, since the value of 
$\alpha^{\s R}_{s}(\mu^2)$ is adjusted to 
compensate precisely the $\mu$-dependence 
in the other components.

\begin{figure}[t!]
\includegraphics[width=0.6\linewidth]{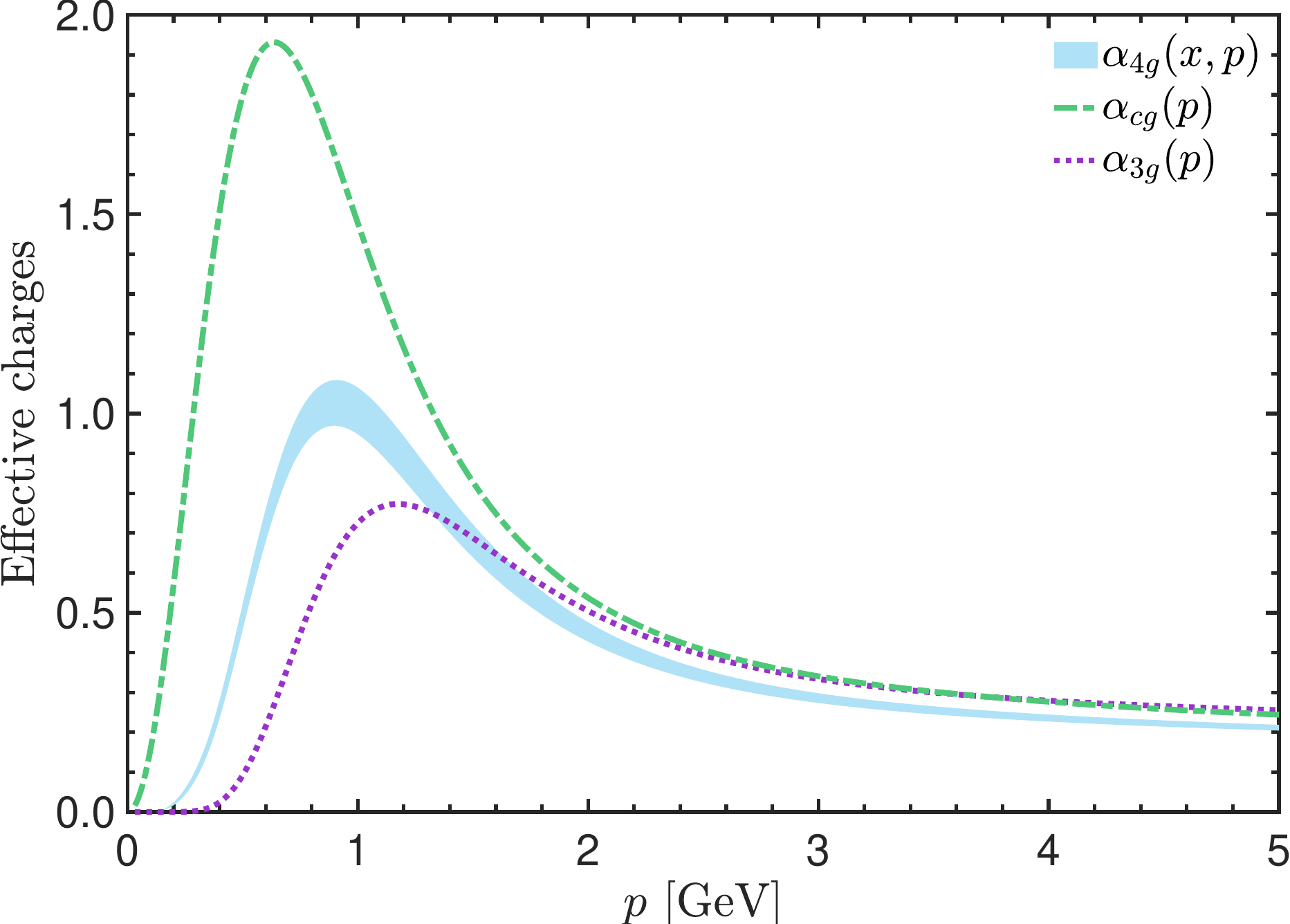}
\caption{ The effective charges $\alpha_{\srm{4g}}(x,p)$, $\alpha_{\srm{cg}}(p)$,  and $\alpha_{\srm{3g}}(p)$, defined in Eqs.~\eqref{4geff} and \eqref{coups4g_3g}, respectively.}
\label{fig:Charges}
\end{figure}

In order to obtain a concrete realization of 
a vertex-derived effective charge, one normally specifies a particular kinematic configuration for
the vertex form factor entering in it.
For instance, in the case of the effective charges 
$\alpha_{\srm{cg}}(p)$ and 
$\alpha_{\srm{3g}}(p)$,  
obtained from the ghost-gluon and three-gluon vertices, 
respectively\cite{Alkofer:2004it,Kellermann:2008iw,Aguilar:2009nf,Aguilar:2010gm,Boucaud:2011ug}, 
\be
\alpha_{\srm{cg}}(p) =\, \alpha_s(\mu^2) B_{sg}^2(p^2)F^2(p^2){\cal Z}(p^2) \,, 
\qquad
\alpha_{\srm{3g}}(p) =\, \alpha_s(\mu^2) \Ls^2(p^2){\cal Z}^3(p^2) \,,
\label{coups4g_3g}
\ee
where $B_{sg}(p^2):=B_1(p,-p,0)$ and $\Ls(p^2)$ are the ``soft-gluon'' limits of the classical form factors of the ghost-gluon and three-gluon vertices, respectively. Instead, 
in order to obtain a more representative 
picture for the  strength of the 
four-gluon interaction, 
we evaluate 
$\alpha_{\srm{4g}}(x,p)$ using  
not one, but all fourteen collinear 
configurations listed 
in \1eq{random_kin}. Thus, we generate  
a sequence of effective charges, which end up 
comprising the blue band shown in 
\fig{fig:Charges}.

For comparison, in the same figure 
we show also the ghost-gluon and three-gluon effective charges of \1eq{coups4g_3g}. For momenta $p\lessapprox1.5~\text{GeV}$ there is a clear hierarchy,  which is expressed by the inequality $\alpha_{\srm{3g}}(p) < \alpha_{\srm{4g}}(x,p) < 
\alpha_{\srm{cg}}(p)$, independently of the collinear configuration chosen for $\alpha_{\srm{4g}}(x,p)$. In fact, the same hierarchy was obtained in~\cite{Cyrol:2014kca,Huber:2018ned} for charges defined by the three- and four-gluon vertices in their totally symmetric configurations, and the soft-ghost kinematics for the ghost-gluon vertex.

In the ultraviolet, the different charges slowly merge into each order. Nevertheless, even at $p = 5$~GeV, the $\alpha_{\srm{4g}}$, $\alpha_{\srm{3g}}$ and $\alpha_{\srm{cg}}$ are visibly different. Moreover, for $p\gtrapprox 3.5$~GeV, the charge hierarchy is modified
with respect to the infrared: $\alpha_{\srm{4g}}(x,p) < 
\alpha_{\srm{cg}}(p) < \alpha_{\srm{3g}}(p)$. To verify that this ordering and the overall sizes of the charges in the ultraviolet are correct, rather than a numerical or truncation artifact, we compute the relation between these charges at one loop.

Perturbatively, one charge can be expanded in powers of any other~\cite{Celmaster:1979km}; in particular, taking $x = x_0$
\be 
\alpha{\srm{3g}}(p) = \alpha_{\srm{4g}}(x_0,p) + a_1\alpha_{\srm{4g}}^2(x_0,p) + \ldots\,, \qquad  \alpha_{\srm{cg}}(p) = \alpha_{\srm{4g}}(x_0,p) + b_1\alpha_{\srm{4g}}^2(x_0,p) + \ldots \,, \label{charge_exp}
\ee
where the ellipses denote higher powers of $\alpha_{\srm{4g}}$. Then, our task reduces to determining the one-loop coefficients $a_1$ and $b_1$. Using the results of~\cite{Pascual:1984zb} for $B_{sg}(p^{2})$, $F(p^{2})$, ${\cal Z}(p^{2})$, and $\Ls(p^{2})$, together with the one-loop result for $G_1(x_0,p)$ given in \1eq{G1_1loop},
we find that
\begin{align}
    a_{1} = \frac{1}{96\pi}\left[ 161 + 28\ln(2)\right] \approx 0.60\,, &&
    b_{1} = \frac{1}{96\pi}\left[47 + 28\ln(2)\right] \approx 0.22\,.
\end{align}
Since $a_1 > b_1 > 0$, \1eq{charge_exp} implies that the hierarchy found in \fig{fig:Charges} at large $p$ agrees with perturbation theory. 
In fact, it is straightforward to 
confirm that the agreement 
is not just qualitative, but 
quantitative. In particular, 
setting  
$p = 4.3$~GeV and $\alpha_{\srm{4g}}= 0.23$ [see \1eq{scheme_conversion_fin}] 
into \1eq{charge_exp}, we get 
$\alpha_{\srm{cg}}= 0.24$, and $\alpha_{\srm{3g}}= 0.26$; these 
values match 
those obtained from the full nonperturbative calculation 
(\ie the corresponding curves in \fig{fig:Charges}), namely 
$\alpha_{\srm{cg}}= 0.26$ and $\alpha_{\srm{3g}}= 0.27$, within 
$8\%$ and $3\%$, respectively.

\section{Conclusions}
\label{conc}

We have presented a comprehensive 
nonperturbative study of the 
transversely-projected 
four-gluon vertex by means of the SDE obtained within the 4PI effective action formalism. 
For the purposes of this work we have restricted our attention to the case of the collinear kinematics, 
which simplify considerably the 
complexity associated with this vertex. This kinematic choice, 
in conjunction with the elimination 
of color terms of the type $f^{abx}
d^{cdx}$ 
by virtue of the charge conjugation symmetry, reduces the number of 
possible form factors down to fifteen. A special subset of three 
form factors, including the one 
associated with the tree-level 
(classical) tensorial structure, 
is subsequently singled out and studied in detail. 
Our treatment of the SDE  
uses as 
external inputs results obtained 
from lattice QCD and previous 
continuous studies.

The most outstanding nonperturbative 
feature that emerges from 
our analysis is the ``planar degeneracy'' of the 
form factors considered.  
Specifically, at a high 
level of accuracy, all kinematic 
configurations that lie on a 
given plane share the same form factors in the infrared. This special property, first observed at the level 
of the three-gluon vertex~\mbox{\cite{
Eichmann:2014xya,Ferreira:2023fva,
Pinto-Gomez:2022brg,Aguilar:2023qqd}}, 
appears to be shared by the 
four-gluon vertex, at least 
in the context of the collinear configurations that we have studied. 
It would be clearly 
important to probe the extent of validity of 
the planar degeneracy for 
more complicated 
configurations of momenta, 
such as general soft-gluon, where one momentum vanishes and the other three are completely arbitrary;
calculations in this direction are already in progress.

It would be of course particularly interesting to confront our findings 
with lattice QCD, as has been the 
case with all other 
fundamental vertices of the theory. 
Unfortunately, the 
preliminary data 
available~\cite{Catumba:2021qbh,Colaco:2023qin,Colaco:2024gmt}
are not sufficiently conclusive 
for confirming any 
of the main features 
encountered in the present study. 
It would be interesting to refine the lattice simulations, in order for a meaningful comparison with the 
results of functional methods to 
become feasible. In this context, the assumption of planar degeneracy might prove particularly useful, 
because it would increase
significantly the available statistics.  

Let us finally point out 
that in our analysis 
we have not considered
dynamical quarks. The 
inclusion of quarks may proceed straightforwardly, 
by making two basic modifications to the treatment  
presented here. First, 
in the diagrammatic representation of the SDE 
in \fig{fig:SDE4g},
one has to include 
additional diagrams containing quark-loops~\cite{Oliveira:2018lln,Kizilersu:2021jen}. Second, the inputs employed, such as 
gluon and ghost propagators, as well as 
ghost-gluon and three-gluon vertices, 
must be replaced by their ``unquenched'' counterparts, 
see~\cite{Ayala:2012pb,Binosi:2016xxu,Aguilar:2019uob}, 
~\cite{Ilgenfritz:2006he,Williams:2015cvx,Aguilar:2023mam},
and~\cite{Aguilar:2019uob}, respectively.
We hope to be able to present a
study of the four-gluon vertex with $2+1$ quark flavors 
in the near future.

\section{Acknowledgments}
\label{sec:acknowledgments}

We thank F.~De~Soto, O.~Oliveira, J.~Rodr\'iguez-Quintero, and  P.~J.~Silva for several useful interactions.  The work of  A.~C.~A. and L.~R.~S. are supported by the CNPq grants \mbox{307854/2019-1} and 
\mbox{162264/2022-4}.
A.~C.~A also acknowledges financial support from project 464898/2014-5 (INCT-FNA).
M.~N.~F. and J.~P. are supported by the Spanish MICINN grant PID2020-113334GB-I00. M.~N.~F. acknowledges financial support from Generalitat Valenciana through contract \mbox{CIAPOS/2021/74}. J.~P. also acknowledges funding from the Generalitat Valenciana grant CIPROM/2022/66.

\appendix

\section{Bose symmetric basis for collinear configurations}
\label{basis_col}

In this appendix we briefly describe the $S_4$ permutation group formalism employed for the construction of the tensor basis of \1eq{basis} for $\overline{\fatg}^{abcd}_{\mu\nu\rho\sigma}(x,p)$. A detailed account of this method in the context of four-point functions is given in~\cite{Eichmann:2015nra}.

The primary objective of  this construction is to obtain tensors $t_{i,\mu\nu\rho\sigma}^{abcd} (x,p)$ that are invariant under a permutation of any two external legs of the vertex. To accomplish this, one first identifies multiplets of the $S_4$ permutation group, which transform nontrivially under permutations (see Eq.~(47) of ~\cite{Eichmann:2015nra}). Then, one can construct singlets (invariants) by taking appropriately defined products of these multiplets.

We begin by rearranging the momenta $p_i$ into the Mandelstam momenta, which for collinear configurations read
\be 
q := p_1 + p_3 = ( x_1 + x_3 )p\,, \qquad t := p_2 + p_3 = ( x_2 + x_3 )p \,, \qquad k := p_1 + p_2 = ( x_1 + x_2 )p \,.
\ee
From these momenta, we form six scalars which are chosen to be
\begin{align}
    q^2 =& ( x_1 + x_3 )^2 p^2 \,, \quad & w_{1} :=& q\cdot k = ( x_1 + x_3 )( x_1 + x_2 )p^2\,,\nonumber\\ 
    k^2 =& ( x_1 + x_2 )^2 p^2 \,, \quad &w_{2} :=& t\cdot k = ( x_2 + x_3 )( x_1 + x_2 )p^2\,, \nonumber\\
    t^2 =& ( x_2 + x_3 )^2 p^2 \,, \quad  &w_{3} :=& q\cdot t = ( x_1 + x_3 )( x_2 + x_3 )p^2\,. \label{scalars}
\end{align}

The above scalars can then be combined to form an $S_4$ singlet, 
\begin{align}\label{scalar_sing}
    \sb^{2} = \frac{1}{2}(t^{2} + q^{2} + k^{2}) = \frac{p^{2}}{2}\left[x_{1}^{2}+x_{2}^{2}+x_{3}^{2}+x_{4}^{2}\right] \,,
\end{align}
together with a doublet, $\overline{\mathcal{D}}$, and a triplet, $\overline{\mathcal{T}}$, given by
\begin{align}\label{scalar_mult}
    \overline{\mathcal{D}} = \frac{1}{2 \sb^{2}}
    \begin{bmatrix}
        \sqrt{3}(q^{2}-t^{2})\\
        t^{2}+q^{2}-2k^{2}
    \end{bmatrix}\,, && 
    \overline{\mathcal{T}} = \frac{1}{8 \sb^{2}}
    \begin{bmatrix}
        2(w_{1} + w_{2} + w_{3})\\
        \sqrt{2}(w_{1} + w_{2} -2w_{3})\\
        \sqrt{6}(w_{2} - w_{1})
    \end{bmatrix}\,.
\end{align}
Note that the denominator $\sb^2$ in \1eq{scalar_mult} is introduced such that $\overline{\mathcal{D}}$ and $\overline{\mathcal{T}}$ are dimensionless.

Next, the three Lorenz tensors in \1eq{LorTens} can be organized into a singlet, $\mathcal{S}_{L}$, and a doublet, $\mathcal{D}_{L}$, given respectively by
\begin{align}\label{Lorents_mult}
    \mathcal{S}_{L} = P_{\mu\sigma}P_{\nu\rho} + P_{\mu\rho}P_{\nu\sigma} + P_{\mu\nu}P_{\rho\sigma}\,, &&
    \mathcal{D}_{L} = 
    \begin{bmatrix}
        \sqrt{3}(P_{\mu\rho} P_{\nu\sigma}-P_{\mu\sigma}P_{\nu\rho}) \\
        P_{\mu\sigma}P_{\nu\rho} + P_{\mu\rho}P_{\nu\sigma} -2 P_{\mu\nu} P_{\rho\sigma}
    \end{bmatrix}\,.
\end{align}
Similarly, the five color tensors of \1eq{colorTens} may be collected into a singlet,
\be 
\mathcal{S}_{C} = \delta^{ab} \delta^{cd}+\delta^{ac} \delta^{bd}+\delta^{ad} \delta^{bc} \,, \label{color_sing}
\ee
and two doublets,
\begin{align}\label{color_mult}
    \mathcal{D}_{C}^{(1)} = 
    \frac{1}{2}\begin{bmatrix}
        \sqrt{3} (\delta^{ac} \delta^{bd}-\delta^{ad} \delta^{bc})\\
        (-2 \delta^{ab} \delta^{cd}+\delta^{ac} \delta^{bd}+\delta^{ad} \delta^{bc})
    \end{bmatrix}\,,&&
    \mathcal{D}_{C}^{(2)} = 
    \frac{1}{2}\begin{bmatrix}
    \frac{1}{ \sqrt{3}}(f^{cae} f^{bde}-2 f^{abe} f^{cde}+f^{ade} f^{bce})\\
     (f^{ade} f^{bce}-f^{cae} f^{bde})
    \end{bmatrix}\,.
\end{align}

At this point, we define special products of $S_4$ multiplets that will be used to construct invariants. To this end, let ${\cal S}$ denote a generic singlet, and
\begin{align}
    \mathcal{D} = 
    \begin{bmatrix}
        a\\
        s
    \end{bmatrix}\,, && 
    \mathcal{T} = 
    \begin{bmatrix}
        u\\
        v\\
        w
    \end{bmatrix}\,,
\end{align}
denote generic doublets and triplets, respectively. Then we define the dot (``$\cdot$'') product as
\be 
\mathcal{D}\cdot \mathcal{D}^\prime =a a^\prime + s s^\prime \,, \qquad \mathcal{T}\cdot \mathcal{T}^\prime = uu' + vv' + ww' \,,
\ee
together with ``star'' (``$\star$'') and ``wedge'' (``$\wedge$'') products given by
\begin{align}
    \mathcal{D}*\mathcal{D}' = 
    \begin{bmatrix}
        a s' + s a'\\
        a a' - s s'
    \end{bmatrix}\,, && 
        \mathcal{T}*\mathcal{T}' = 
    \begin{bmatrix}
        v w' + w v' + \sqrt{2}(u w' + wu')\\
        w w' - v v' + \sqrt{2}( u v' + v u')
    \end{bmatrix}\,, 
\end{align}
and
\begin{align}
    \mathcal{T}\wedge \mathcal{D} = 
    \begin{bmatrix}
        va-ws\\
        ua - \frac{1}{\sqrt{2}}(va + ws)\\
        -us -  \frac{1}{\sqrt{2}}(vs-wa)
    \end{bmatrix}\,, && 
        \mathcal{D}\wedge \mathcal{D}' = as' - sa'\,.
\end{align}

Using the above products, it can be shown that combinations of the types 
\begin{align} 
&{\cal S} {\cal S}^\prime\,,  \qquad \mathcal{D}\cdot\mathcal{D} \,,  \qquad \mathcal{T}\cdot\mathcal{T} \,, \qquad \mathcal{D}\cdot(\mathcal{D}^\prime*\mathcal{D}^{\prime\prime})\,, \nonumber\\
&(\mathcal{T}*\mathcal{T}^\prime)\cdot\mathcal{D} \,,
 \qquad (\mathcal{T}*\mathcal{T}^\prime)\cdot(\mathcal{D}*\mathcal{D}^\prime) \,, \qquad \left[\mathcal{T}\cdot(\mathcal{T}^\prime\wedge\mathcal{D})\right](\mathcal{D}^\prime\wedge\mathcal{D}^{\prime\prime}) \,, &  \label{singlets}
\end{align}
are $S_4$ singlets~\cite{Eichmann:2015nra}. Hence, to construct a Bose symmetric tensor basis for collinear configurations, we form all possible combinations of the above types using the multiplets in Eqs.~\eqref{scalar_sing} through \eqref{color_mult}, and select from these 15 linearly independent singlets.

From the procedure outlined above, we obtain the following basis
\begin{align}
      t_{1} &= - \mathcal{D}_{L}\cdot \mathcal{D}_{C}^{(2)}\,, 
    & t_{6} &= \mathcal{S}_{C} (\overline{\mathcal{D}} \cdot \mathcal{D}_{L})\,,
    & t_{11}&= (\overline{\mathcal{T}}*\overline{\mathcal{T}})\cdot (\mathcal{D}_{L}*\mathcal{D}_{C}^{(1)})\nonumber\\
      t_{2} &= - \mathcal{D}_{L}\cdot \mathcal{D}_{C}^{(1)}\,,
    & t_{7} &= \mathcal{S}_{L} [(\overline{\mathcal{T}}*\overline{\mathcal{T}})\cdot \mathcal{D}_{C}^{(1)}]\,,
    & t_{12}&= (\overline{\mathcal{T}}*\overline{\mathcal{T}})\cdot (\mathcal{D}_{L}*\mathcal{D}_{C}^{(2)})\,,\nonumber\\
      t_{3} &= - \mathcal{S}_{L} \mathcal{S}_{C}\,,
    & t_{8} &= \mathcal{S}_{L} [(\overline{\mathcal{T}}*\overline{\mathcal{T}})\cdot \mathcal{D}_{C}^{(2)}]\,,
    & t_{13}&= \mathcal{S}_{C} [(\overline{\mathcal{T}}*\overline{\mathcal{T}})\cdot \mathcal{D}_{L}] \,, \label{tibasis}\\
      t_{4} &= - \mathcal{S}_{L} (\overline{\mathcal{D}} \cdot \mathcal{D}_{C}^{(1)})\,,
    & t_{9} &= \overline{\mathcal{D}}\cdot(\mathcal{D}_{L}*\mathcal{D}_{C}^{(1)})\,,
    & t_{14}&= [\overline{\mathcal{T}}\cdot(\overline{\mathcal{T}}\wedge \overline{\mathcal{D}})](\mathcal{D}_{L}\wedge \mathcal{D}_{C}^{(1)})\,,\nonumber\\
      t_{5} &= \mathcal{S}_{L} (\overline{\mathcal{D}} \cdot \mathcal{D}_{C}^{(2)})\,, 
    & t_{10}&= \overline{\mathcal{D}}\cdot(\mathcal{D}_{L}*\mathcal{D}_{C}^{(2)})\,,
    & t_{15}&= [\overline{\mathcal{T}}\cdot(\overline{\mathcal{T}}\wedge \overline{\mathcal{D}})](\mathcal{D}_{L}\wedge \mathcal{D}_{C}^{(2)})\,.\nonumber
\end{align}
Note that the dependence on the $x_i$ appears only through the scalar multiplets, $\sb$, $\overline{\mathcal D}$ and $\overline{\mathcal T}$, of \2eqs{scalar_sing}{scalar_mult}. Hence, the tensors $t_{1,2,3}$ are independent of $x_i$, whereas all others depend on it.

\section{Changing renormalization schemes}
\label{scheme}

The data for the three-gluon and ghost-gluon form factors, 
$\Ls(s^2)$ and $B_1(q,r,p)$, respectively,  
used as external inputs in our SDE, were originally renormalized in a scheme different than that of \2eqs{MOM}{ren_conds_4g} 
that we employ in our analysis. Specifically, $\Ls(s^2)$ and $B_1(q,r,p)$ were computed in~\cite{Aguilar:2022thg,Ferreira:2023fva} in the ``asymmetric MOM scheme'', defined by
\be 
\Delta^\asym(\mu^2) = \mu^{-2} \,, \qquad F^\asym(\mu^2) = 1 \,, \qquad \Ls^\asym(\mu^2) = 1 \,, \label{asym}
\ee
where we introduce the superscript ``$\rm{asym}$'' to denote quantities renormalized in this scheme; quantities without an index specifying a renormalization scheme, such as $\Delta(q^2)$ and $\Ls(r^2)$, will be understood to be renormalized according to \2eqs{MOM}{ren_conds_4g}. 

Note that for the gluon and ghost propagators the renormalization prescriptions of \2eqs{asym}{MOM} coincide. Therefore, the multiplicative renormalizability expressed by \1eq{renconst} implies that the renormalization constants $Z_A$ and $Z_c$ have the same values in both schemes,  \ie
\be
Z_{A} = Z_{A}^\asym = \mu^2 \Delta_{\srm B}(\mu^2) \,, \qquad Z_c = Z_c^\asym = F_{\srm B}(\mu^2) \,, \label{ZA_Zc}
\ee
where the index ``${\rm B}$'' denotes unrenormalized quantities\footnote{In the main text, unrenormalized quantities are not tagged with any index. Here, since quantities renormalized in two different schemes coexist with unrenormalized ones, we introduce the index ``${\rm B}$'' for clarity.}. It follows that \mbox{$\Delta^\asym(q^2) = \Delta(q^2)$} and $F^\asym(q^2) = F(q^2)$.

On the other hand, the vertex form factors and the value of the coupling are different in the two schemes considered. Specifically, \1eq{renconst} implies
\begin{align}
    \Ls(r^2) = \frac{Z_{3}}{\Zthreeasym} \Ls^\asym(r^{2})\,, &&
    B_{1}(q,r,p) = \frac{Z_{1}}{\Zoneasym} B_{1}^\asym(q,r,p)\,, && 
    \alpha_{s}= \left(\frac{\Zgasym}{Z_{g}}\right)^{\!\!\!2}\!\alphaasym \,.
    \label{scheme_conversion}
\end{align}
Then, it is easily shown from the STIs of \1eq{eq:sti_renorm} that $Z_{3} = (Z_{4}Z_{A})^{1/2}$, $Z_{1} = Z_{3}Z_{c}Z_{A}^{-1}$, and $Z_{g} = Z_{1}Z_{c}^{-1}Z_{A}^{-1/2}$, and identical relations with ``asym'' superscripts in all quantities. Hence, using \1eq{ZA_Zc}, the ratios in \1eq{scheme_conversion} can be expressed as
\begin{align}
    \frac{Z_{3}}{\Zthreeasym} = \frac{Z_{1}}{\Zoneasym} = \frac{Z_{g}}{\Zgasym} = \left(\frac{Z_{4}}{\Zfourasym}\right)^{\!\!1/2}\,. \label{Zs_ratios}
\end{align}
Thus, in order to convert between the asymmetric MOM and the prescription of \2eqs{MOM}{ren_conds_4g}, we simply need to obtain the \emph{finite}, and regulator independent, ratio $\Zfourasym/Z_{4}$. 

In what follows, we assume that the renormalization point $\mu = 4.3$~GeV is sufficiently large such that $Z_{4}/Z_4^\asym$ can be approximated by a one-loop calculation.

We begin by computing the one-loop $Z_{4}$ in the renormalization scheme of \1eq{ren_conds_4g}. To that end, we substitute in \1eq{eqG1} the tree-level forms of all inputs, \ie $\Delta(q^2)\to 1/q^2$, $\Ls(r^2) \to 1$ and \mbox{$G_1^{*} \to 1$}. Using dimensional regularization with $d = 4 - 2\epsilon$ space-time dimensions, this procedure yields
\be 
G_1^{(1)}(0,p,p,-2p) = Z_4 - \frac{\alpha_s}{96\pi}\left\lbrace 48\left[ \frac{1}{\epsilon} - \gamma_{\srm E} + \ln(4\pi) - \ln\left(\frac{p^2}{\nu^2}\right)\right] +  28 \ln(2) - 11 \right\rbrace \,,
\ee
where $\gamma_{\srm E}$ is the Euler-Mascheroni constant, and $\nu$ is the t'Hooft mass.

Then, the renormalization condition of \1eq{ren_conds_4g} determines $Z_{4}$ as
\begin{align}
   Z_{4} = 1 + \frac{\alpha_{s}}{96 \pi } \left\lbrace 48\left[ \frac{1}{\epsilon } - \gamma_{\srm E} + \ln ( 4 \pi ) - \ln\left(\frac{\mu^2}{\nu^2}\right) \right] + 28 \ln (2) - 11 \right\rbrace\,, \label{Z4_4g}
\end{align}
and finally
\begin{align}
   G_{1}^{(1)}(0,p,p,-2p) &= 1+\frac{\alpha_{s}}{2 \pi } \ln\left(\frac{p^{2}}{\mu^{2}}\right)\,. \label{G1_1loop}
\end{align}

Next, $\Zfourasym$ can be determined by invoking the STIs of \1eq{eq:sti_renorm}, which together imply $Z_4^\asym = (Z_3^\asym)^2 /Z_A^\asym$. Using textbook results~\cite{Pascual:1984zb} for $Z_A^\asym$ and $Z_3^\asym$ entails
\begin{align}
    \Zfourasym = 1 + \frac{ \alphaasym}{24 \pi}\left\lbrace 12 \left[\frac{1}{\epsilon } - \gamma_{\srm E} + \ln (4 \pi ) - \ln\left(\frac{\mu^2}{\nu^2}\right)  \right] - 43\right\rbrace\,. \label{Z4_asym}
\end{align}

At this point, \2eqs{Z4_4g}{Z4_asym} involve couplings, $\alphaasym$ and $\alpha_s$, in two different renormalization schemes. In the perturbative framework of the present calculation, we can expand one coupling in a power series of the other~\cite{Celmaster:1979km}, \ie $\alpha_{s} = \alpha_{s}^\asym + {\cal O}((\alpha_{s}^\asym)^2)$. Then, at first order in $\alphaasym$
\begin{align}
\left(\frac{Z_{4}}{\Zfourasym}\right)^{\!\!1/2} = 1 + \frac{7\alpha_s^\asym}{192\pi}\left[ 23 + 4\ln(2) \right] \,. \label{Z4_ratio}
\end{align}

Finally, using the value of $\alphaasym(\mu^2) = 0.27$, for $\mu = 4.3$~GeV, determined in~\cite{Boucaud:2017obn}, we obtain
\begin{align}
\left(\frac{Z_{4}}{\Zfourasym}\right)^{\!\!1/2}\approx 1.08\,.
\end{align}
Combining \2eqs{Z4_ratio}{Zs_ratios} into \1eq{scheme_conversion} we obtain the final conversion factors and value of $\alpha_s$,
\be
\Ls(r^2) = 1.08 \, \Ls^\asym(r^{2})\,, \qquad
    B_{1}(q,r,p) = 1.08 \, B_{1}^\asym(q,r,p)\,, \qquad \alpha_{s}(\mu^2) = 0.23 \,,
    \label{scheme_conversion_fin}
\ee
valid for $\mu = 4.3$~GeV.


%

\end{document}